\documentclass[a4paper,11pt]{article}
\pdfoutput=1 

\usepackage{jcappub} 

\usepackage[T1]{fontenc} 

\usepackage{empheq}

\usepackage{listings}

\usepackage{tablefootnote}

\title{\boldmath Analytical Template for the 4-Point Correlation Function Covariance Beyond the Gaussian Random Field ${\rm I}$: 1-Loop Corrections involving Second-Order Densities}
\author[1]{William Ortolá Leonard,} 
\author[2]{Zachary Slepian}

\affiliation[1]{Department of Physics, University of Florida,\\2001 Museum Rd., Gainesville, FL 32611, USA}
\affiliation[2]{Department of Astronomy, University of Florida,\\211 Bryant Space Science Center, Gainesville, FL 32611, USA}

\emailAdd{wortola@ufl.edu}
\emailAdd{zslepian@ufl.edu}

\abstract{Analytical templates for the covariance matrix of the 4-Point Correlation Function (4PCF) have been developed in the past assuming a Gaussian Random Field (GRF). In this work, we present the first non-Gaussian calculation of the 4PCF covariance, incorporating 1-loop corrections using the second-order density contrast. Furthermore, we introduce a non-trivial galaxy bias scheme at second order. To simplify the calculation, we decompose the covariance into five distinct structures, and then exploit the isotropic basis functions of \cite{Cahn_Iso}. This approach reduces the complexity of the high-dimensional integrals naively involved, enabling the angular parts to be performed and leaving us with low-dimensional radial integrals. This analytical template will provide a more accurate characterization of the statistical errors on the 4PCF, improving both the parity-odd and parity-even analyses. This is the first paper in a two-part series.}

\usepackage{empheq}
\usepackage{listings}
\usepackage{hyperref}
\usepackage{tablefootnote}
\usepackage{graphicx} 

\newcommand{\vecx}{\mathbf{x}}
\newcommand{\vecr}{\mathbf{r}}
\newcommand{\vecs}{\mathbf{s}}
\newcommand{\veck}{\mathbf{k}}
\newcommand{\vecq}{\mathbf{q}}
\newcommand{\dg}{\delta_{\rm g}}
\newcommand{\dD}{\delta_{\rm D}^{[3]}}
\newcommand{\dK}{\delta^{\rm K}}
\newcommand{\dlin}{\delta_{\rm lin.}}
\newcommand*\widefbox[1]{\fbox{\hspace{1em}#1\hspace{1em}}}
\newcommand{\hatk}{\widehat{\mathbf{k}}}
\newcommand{\hats}{\widehat{\mathbf{s}}}
\newcommand{\hatr}{\widehat{\mathbf{r}}}
\newcommand{\hatq}{\widehat{\mathbf{q}}}
\newcommand{\hatn}{\widehat{\mathbf{n}}}
\newcommand{\hatR}{\widehat{\mathbf{R}}}
\newcommand{\Pk}{P_{\rm lin}}
\newcommand{\PP}{\mathcal{P}}
\newcommand{\W}{W^{(2)}}

\newcommand{\tj}[6]{\begin{pmatrix} {#1} & {#2} & {#3}\\ {#4} & {#5} & {#6}\end{pmatrix}}

\begin{document}
\maketitle
\flushbottom

\section{Introduction}
The Universe's Large-Scale Structure (LSS) provides a unique laboratory for studying nature's fundamental symmetries. A powerful probe of these is the 4-Point Correlation Function (4PCF), which enables both cosmological inference and searches for parity-violating signals. The connected, even- and odd-parity 4PCF have been measured using data from the Sloan Digital Sky Survey (SDSS) Baryon Oscillation Spectroscopic Survey (BOSS) by \cite{Philcox_4PCF_measurement, Hou_Parity, Philcox_Parity}; notably, \cite{Hou_Parity} reported the first detection of parity-odd modes at $7.1\sigma$, followed by \cite{Philcox_Parity} with $2.9\sigma$ evidence. More recently, using data from Year 1 (Y1) of the Dark Energy Spectroscopic Instrument (DESI), \cite{Hou_4PCF_Measurement} measured the connected even-parity 4PCF, while \cite{Slepian_Parity_odd} found evidence for parity-violating signals at the $4$–$10\sigma$ level. These statistical significances are determined through the covariance matrix, which plays a central role in quantifying the uncertainties and correlations of the measured 4PCF coefficients in the isotropic basis of \cite{Cahn_Iso, Iso_gen}. These analyses rely on the formalism introduced in \cite{cahn_prl} and adopt covariance matrices constructed under the Gaussian approximation, as in \cite{Hou_Cov}.

In this approach, the 4PCF covariance is obtained by applying Wick’s theorem, reducing the eight-point function into products of two-point functions and thus neglecting connected higher-order contributions. While this Gaussian covariance captures the dominant structure of the matrix and enables practical analyses, it omits non-Gaussian effects arising from nonlinear mode coupling in the density field, and from beyond-linear galaxy bias. Assessing the impact of these omitted contributions on inferred uncertainties, and hence on reported detection significances, requires going beyond the Gaussian approximation. These contributions are expected to become increasingly important on mildly non-linear scales.

Previous works have computed power spectrum covariance matrices beyond the Gaussian Random Field (GRF) approximation for both matter \cite{Bertolini_Cov, Mohammed_Cov} and galaxies \cite{Wadekar_Cov}. For the 4PCF, however, the covariance has only been modeled assuming a GRF density \cite{Hou_Cov}, as mentioned above. This motivates a systematic computation of non-Gaussian corrections to the 4PCF covariance and an assessment of their relevance for current analyses. In this work, we compute the next-to-leading-order (NLO) contribution to the 4PCF covariance. Our approach is based on Standard Perturbation Theory (SPT), including galaxy bias up to second order, and builds upon the approach of \cite{Ortola_4PCF_Cov_II}; companion paper to this work dealing with third-order contributions. A detailed assessment of the impact of these NLO contributions on parameter inference and detection significances is left for future work.

The structure of this work is as follows. In $\S$\ref{sec:Cov_Compu}, we construct the NLO covariance matrix, corresponding to tenth order in the linear density field. In $\S$\ref{sec:Resulting_Cov}, we present its explicit mathematical form. The appendices contain the detailed derivations and mathematical tools required for these results.

\section{Setting Up the 1-Loop Covariance with Second-Order Densities} \label{sec:Cov_Compu}
The covariance is defined as \cite{Hou_Cov}:
\begin{align}
{\rm Cov}(\zeta(\mathbf{R}),\zeta(\mathbf{R}')) \equiv \left< \zeta(\mathbf{R})\zeta(\mathbf{R}')\right> - \big<\zeta(\mathbf{R})\big> \left<\zeta(\mathbf{R}')\right>,
\end{align}
where $\zeta$ represents any N-Point Correlation Function (NPCF) we wish to evaluate and the angle bracket represents an average over realizations of the density field. We define $\mathbf{R} \equiv (\vecr_0,\vecr_1,\cdots,\vecr_{N-1})$, and $\mathbf{R}'$ in the same way. Following the procedure of \cite{Hou_Cov}, we evaluate the first term:
\begin{align}\label{eq:Con_Cov_First_Expansion}
\left< \zeta(\mathbf{R})\zeta(\mathbf{R}')\right> = \int\frac{ d^3\vecs}{V} \left< \prod_{i=0}^{N-1} \;\dg(\vecx+\vecr_i)\;\dg(\vecx+\vecr'_i+ \vecs) \right>,
\end{align}
where the angle brackets represent the ensemble average of $\vecx$. We set $N=4$, \textit{i.e.,} we evaluate the 4PCF. Subscript ``g'' indicates the density fluctuations, $\delta$, are for galaxies, and $V$ is the survey volume. Using the Eulerian bias expansion \cite{Bernardeau, Scoccimarro, Desjacques_bias_review}:
\begin{align}\label{eq:dc_galaxy_expansion}
\dg(\vecx) = b_0 + b_1\delta_{\rm m}(\vecx) + \frac{b_2}{2}\delta_{\rm m}^{2}(\vecx) \ + b_{S} \;S^{(2)}(\vecx),\footnotemark
\end{align}
\footnotetext{Other works may use $b_2$ in place of our $b_2/2$, \textit{e.g.} page 2 of \cite{Slepian_3PCF}.}with $b_0$ ensuring $\langle\dg\rangle = 0$; however, $b_0$ will be omitted in the rest of this work since it does not enter connected correlation functions \cite{Scoccimarro}. $b_1$ is the \textit{linear bias}, which describes how the galaxy density traces the
matter’s density fluctuation linearly, $b_2$ the \textit{quadratic bias}, which describes how the galaxy density traces the matter's density fluctuation squared, and $b_S$ the tidal tensor bias, which captures how galaxies respond not just to the local density, but to anisotropic stretching and compression of matter (the local tidal field). Then, the matter density contrast is expanded using SPT:
\begin{align}\label{eq:dc_matter_expansion}
\delta_{\rm m}(\vecx) = \delta_{\rm lin.} (\vecx)+ \delta^{(2)}(\vecx) + \mathcal{O}(\delta_{\rm lin.}^3).
\end{align}
Above, we expanded the matter density contrast to second order in the linear density contrast, $\delta_{\rm lin.}$; we define $\delta^{(2)}(\vecx)$, $\delta_{\rm lin.}^{2}(\vecx)$ and $S^{(2)}(\vecx)$ by their inverse Fourier Transforms in Appendix \ref{sec:Generalization_delta_2}. We analyze the third-order terms in the second paper of the series \cite{Ortola_4PCF_Cov_II}. Hence, inserting Eq. (\ref{eq:dc_matter_expansion}) into Eq. (\ref{eq:dc_galaxy_expansion}), and incorporating the resultant expression into Eq. (\ref{eq:Con_Cov_First_Expansion}), we obtain:
\begin{align}\label{eq:Con_Cov_second_Expansion}
&\left< \zeta(\mathbf{R})\zeta(\mathbf{R}')\right> = \int\frac{ d^3\vecs}{V} \Bigg\langle
\prod_{i=0}^{3}
\bigg[ b_1\left(\delta_{\text{lin}}+\delta^{(2)}\right) + \tfrac{b_2}{2}\left(\delta_{\text{lin}}+\delta^{(2)}\right)^2 + b_{S}\, S^{(2)} \bigg](\mathbf{x}+\mathbf{r}_i)\nonumber \\
&\qquad\qquad\qquad\qquad\qquad\times
\bigg[ b_1\left(\delta_{\text{lin}}+\delta^{(2)}\right) + \tfrac{b_2}{2}\left(\delta_{\text{lin}}+\delta^{(2)}\right)^2 + b_{S}\, S^{(2)} \bigg](\mathbf{x}+\mathbf{r}'_{i}+\mathbf{s})\Bigg\rangle.
\end{align}
Here, the arguments $\vecx+\vecr_i$ and $\vecx+\vecr'_i+\vecs$ apply to all terms inside the square brackets. Carrying out the product, we obtain the 1-loop covariance with second-order densities as:
\begin{align}\label{eq:Full_cov}
&{\rm Cov}_{4, {\rm 1L}}^{[2]}(\mathbf{R},\mathbf{R}')=\nonumber \\ 
& \qquad b_1^8  \bigg\{ \sum_{i=0}^{3}\sum_{j=0}^{3} \bigg< \delta^{(2)}(\vecx + \vecr_{i})\delta^{(2)}(\vecx+\vecr'_{j}+\vecs)  \prod_{p=0,p\neq i}^{3} \prod_{t=0,t\neq j}^{3} \delta_{\rm lin.}(\vecx+\vecr_p)\delta_{\rm lin.}(\vecx+\vecr'_t+ \vecs) \bigg>  \bigg\} \nonumber \\ 
& \qquad +  b_1^6 \left(\frac{b_2}{2}\right)^2 \bigg\{\sum_{i=0}^{3}\sum_{j=0}^{3} \bigg< \delta_{\rm lin.}^2(\vecx + \vecr_i)\delta_{\rm lin.}^2(\vecx+\vecr'_j+\vecs) \nonumber \\ 
& \qquad \qquad \times\prod_{p=0,p\neq i}^{3} \prod_{t=0,t\neq j}^{3} \delta_{\rm lin.}(\vecx+\vecr_p)\delta_{\rm lin.}(\vecx+\vecr'_t+ \vecs) \bigg> \bigg\} \nonumber \\ 
& \qquad +  b_1^6\; b_s^2 \bigg\{\sum_{i=0}^{3}\sum_{j=0}^{3} \bigg< S^{(2)}(\vecx + \vecr_i)S^{(2)}(\vecx+\vecr_j'+\vecs) \nonumber \\ 
& \qquad \qquad \times\prod_{p=0,p\neq i}^{3} \prod_{t=0,t\neq j}^{3} \delta_{\rm lin.}(\vecx+\vecr_p)\delta_{\rm lin.}(\vecx+\vecr'_t+ \vecs) \bigg> \bigg\} \nonumber \\ 
& \qquad +  b_1^7 \frac{b_2}{2}  \bigg\{\sum_{i=0}^{3}\sum_{j=0}^{3} \bigg< \delta_{\rm lin.}^2(\vecx + \vecr_i)\delta^{(2)}(\vecx+\vecr_j'+\vecs)\nonumber \\ 
& \qquad \qquad \times \prod_{p=0,p\neq i}^{3} \prod_{t=0,t\neq j}^{3} \delta_{\rm lin.}(\vecx+\vecr_p)\delta_{\rm lin.}(\vecx+\vecr'_t+ \vecs) \bigg> + {\rm symm.} \bigg\} \nonumber \\ 
& \qquad +  b_1^6 \;b_s \frac{b_2}{2}  \bigg\{\sum_{i=0}^{3}\sum_{j=0}^{3} \bigg< \delta_{\rm lin.}^2(\vecx + \vecr_i)S^{(2)}(\vecx+\vecr_j'+\vecs)\nonumber \\ 
& \qquad \qquad \times \prod_{p=0,p\neq i}^{3} \prod_{t=0,t\neq j}^{3} \delta_{\rm lin.}(\vecx+\vecr_p)\delta_{\rm lin.}(\vecx+\vecr'_t+ \vecs) \bigg>  +{\rm symm.} \bigg\} \nonumber \\ 
& \qquad +  b_1^7 b_s \bigg\{\sum_{i=0}^{3}\sum_{j=0}^{3} \bigg< S^{(2)}(\vecx + \vecr_i)\delta^{(2)}(\vecx+\vecr_j'+\vecs)\nonumber \\ 
& \qquad \qquad \times \prod_{p=0,p\neq i}^{3} \prod_{t=0,t\neq j}^{3} \delta_{\rm lin.}(\vecx+\vecr_p)\delta_{\rm lin.}(\vecx+\vecr'_t+ \vecs) \bigg> + {\rm symm.} \bigg\}. 
\end{align}
Above, we used the subscript 4,1L on the left-hand side to denote that we evaluate the 4PCF 1-loop covariance. This equation captures all possible permutations of the terms through the two summations and two products, ensuring that for each choice of $i$, $p \in \{0,1,2,3\}$ must be distinct from $i$, and similarly, for each choice of $j$, $t \in \{0,1,2,3\}$ must be distinct from $j$. In what follows, we display only the product over $p$ and $t$, with the understanding that $p \neq i$ and $t \neq j$ are implicitly enforced. The first three terms are symmetric under the exchange of primed and unprimed vectors, whereas the last three terms are not. To restore the symmetry, an additional ``+ symm.'' has been included in the last three terms; we show an example of how to compute these terms below: 
\begin{align}
& b_1^7 \frac{b_2}{2}  \bigg\{\sum_{i=0}^{3}\sum_{j=0}^{3} \bigg< \delta_{\rm lin.}^2(\vecx + \vecr_i)\delta^{(2)}(\vecx+\vecr_j'+\vecs)\nonumber \\ 
& \qquad \qquad \times \prod_{p=0,p\neq i}^{3} \prod_{t=0,t\neq j}^{3} \delta_{\rm lin.}(\vecx+\vecr_p)\delta_{\rm lin.}(\vecx+\vecr'_t+ \vecs) \bigg> + {\rm symm.} \bigg\} \nonumber \\
&\qquad = b_1^7 \frac{b_2}{2}  \bigg\{\sum_{i=0}^{3}\sum_{j=0}^{3} \bigg< \delta_{\rm lin.}^2(\vecx + \vecr_i)\delta^{(2)}(\vecx+\vecr_j'+\vecs)\nonumber \\ 
& \qquad \qquad \times \prod_{p=0,p\neq i}^{3} \prod_{t=0,t\neq j}^{3} \delta_{\rm lin.}(\vecx+\vecr_p)\delta_{\rm lin.}(\vecx+\vecr'_t+ \vecs) \bigg> \nonumber \\
&\qquad \quad +\sum_{i=0}^{3}\sum_{j=0}^{3} \bigg< \delta^{(2)}(\vecx + \vecr_i)\delta_{\rm lin.}^{2}(\vecx+\vecr_j'+\vecs) \nonumber \\ 
& \qquad \qquad \times\prod_{p=0,p\neq i}^{3} \prod_{t=0,t\neq j}^{3} \delta_{\rm lin.}(\vecx+\vecr_p)\delta_{\rm lin.}(\vecx+\vecr'_t+ \vecs) \bigg> \bigg\}.
\end{align}
We prove in Appendix \ref{sec:Generalization_delta_2} that the six terms in Eq. (\ref{eq:Full_cov}) may all be written in terms of a simple expression, using a generalization of the (Fourier space) $F^{(2)}$ kernel that we develop there. 

Using the result of Appendix \ref{sec:Generalization_delta_2} and rewriting the density contrast in terms of its inverse Fourier transform, the 1-loop covariance is:
\begin{align}\label{eq:Cov_pre_expansion}
&{\rm Cov}_{4, \rm 1L}^{[2]}(\mathbf{R},\mathbf{R}') = \prod_{v=0}^{3}\int_{\vecs}\int_{\;\veck_v}\int_{ \; \veck'_v}\left[ \;e^{-i\veck_v\cdot(\vecx+\vecr_v)} e^{-i\veck'_v \cdot(\vecx+\vecr'_v+\vecs)} \right]\nonumber \\
& \qquad \qquad \qquad \times \sum_{\mu,\beta} \sum_{i,j}\left< \widetilde{\delta}_\mu^{(2)}(\veck_i)\widetilde{\delta}_\beta^{(2)}(\veck_j')\prod_{p,t}\widetilde{\delta}_{\rm lin.}(\veck_p) \widetilde{\delta}_{\rm lin.}(\veck'_t)\right>,
\end{align}
where $\mu$ and $\beta$ run over $\{0,1,2\}$ such that $ \left\{\widetilde{\delta}_0^{(2)},\widetilde{\delta}_1^{(2)},\widetilde{\delta}_2^{(2)}\right\} = \left\{S^{(2)}, \widetilde{\delta}^{(2)}, \widetilde{\delta}_{\rm lin.}^2\right\}$, respectively. We have abbreviated the 3D integrals as:
\begin{align}
&\int_{\vecs} = \int \frac{d^{3}\vecs}{V},\;\; \int_{\veck} = \int \frac{d^{3}\veck}{(2\pi)^3}\;\; {\rm and} \;\; \int_{\hatk} = \int d\hatk.\nonumber 
\end{align}

\begin{figure}[h!]
\centering
\includegraphics[scale=0.2]{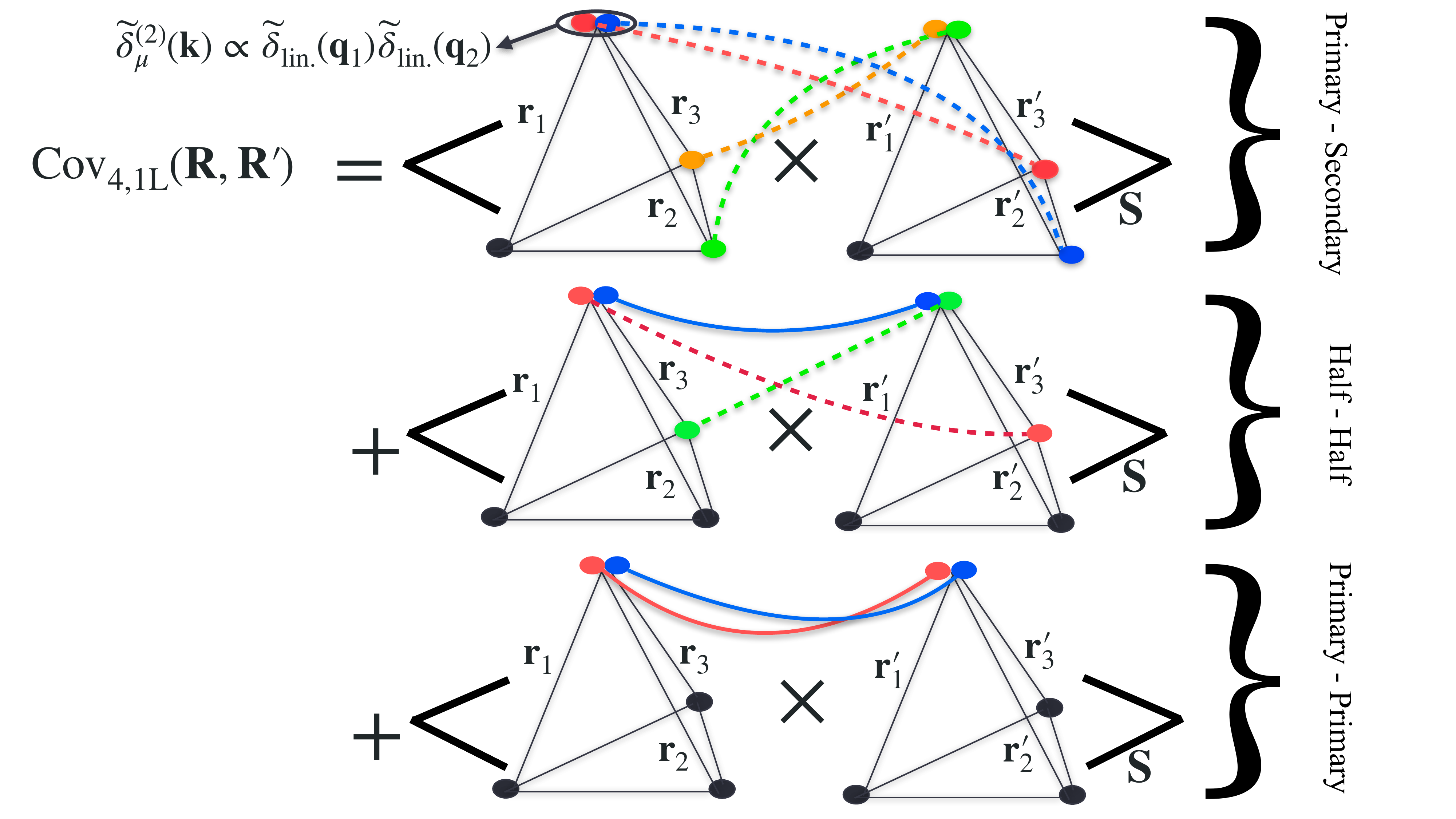}
\caption{Illustration of the three different possible configurations of the fully-coupled covariance at 1-loop order. The three  configurations at this order are: Primary-Secondary, Half-Half and Primary-Primary. The solid lines represent the contraction of two primary density contrasts via Wick's theorem; the dashed lines represent the contraction of a primary overdensity with a secondary one. We define the ‘primary’ points as those that contain the second-order densities and depict them in the above diagram with two colored dots. We define the ‘secondary’ points as those that contain the first-order densities and depict them with one black dot. The above diagram does not show Wick contractions between the leftover-secondary densities (black dots), but what results from them has been explicitly included and calculated in each of the sections below. The primary densities within each tetrahedron are $\vecq_1,\;\vecq_2$ from left to right; likewise with the primed tetrahedron. In the main text, We introduce $\vecr_0 = \vecr'_0 = 0$ to restore the symmetry between arguments of the calculations in $\S$\ref{sec:Resulting_Cov}.}
\label{fig:dc_configurations}
\end{figure}

Expanding the second-order density of Eq. (\ref{eq:Cov_pre_expansion}) using Eq. (\ref{eq:delta_i_def}), we obtain:
\begin{align}\label{eq:sub_cov_before_Wicks}
&{\rm Cov}_{4, \rm 1L}^{[2]}(\mathbf{R},\mathbf{R}') = \prod_{v=0}^{3}\int_{\vecs}\int_{\;\veck_v}\int_{ \; \veck'_v}\left[ \;e^{-i\veck_v\cdot(\vecx+\vecr_v)} e^{-i\veck'_v \cdot(\vecx+\vecr'_v+\vecs)} \right]\nonumber \\
& \qquad \qquad  \times \sum_{\mu,\beta} \int d^3\vecq_1 \int d^3\vecq_2 \int d^3\vecq'_1 \int d^3\vecq'_2 \;W_{\mu}^{(2)}(\vecq_1,\vecq_2) \nonumber \\
& \qquad \qquad  \times W_{\beta}^{(2)}(\vecq'_1,\vecq'_2)\; \dD(\veck_0-\vecq_1-\vecq_2)\; \dD(\veck_0'-\vecq'_1-\vecq'_2)\nonumber \\
& \qquad \qquad  \times  \left< \widetilde{\delta}_{\rm lin.}(\vecq_1)\widetilde{\delta}_{\rm lin.}(\vecq_2)\widetilde{\delta}_{\rm lin.}(\vecq'_1)\widetilde{\delta}_{\rm lin.}(\vecq'_2)\prod_{p,t=1}^{3}\widetilde{\delta}_{\rm lin.}(\veck_p) \widetilde{\delta}_{\rm lin.}(\veck'_t)\right>,
\end{align}
where the $W^{(2)}$ kernel is defined in Eq. (\ref{eq:W_2_def}). The evaluation of the ensemble average of the density contrasts using Wick's theorem leads to five different possible configurations as shown in Figs. \ref{fig:dc_configurations} and \ref{fig:2nd_pc_terms}. Therefore, in the next section we develop the basic mathematical elements of the covariance.

\begin{table} [h!]
\centering
\begin{tabular}{ |p{3cm}|p{3cm}| p{7cm}|}
 \hline
 \multicolumn{3}{|c|}{Table of Coefficients} \\
 \hline
 Coefficient & Equation/Section & Definition \\
 \hline
 &&\\
 $C_{L_1,L_2,L_3}$ & \ref{eq:Constant_from_exp} & Coefficient of plane-wave expansion when projected onto the isotropic basis functions. \\
  &&\\
 $\Upsilon_{L_1,L_2,L_3}$ & \ref{eq:Upsilon_def} & Coefficient from the splitting of position-space-vector and wave-vector isotropic basis functions. \\ 
  &&\\
  $c_{j,n}^{(\mu)}$  &  \ref{eq:W_2_def} & $W_{\delta_i}^{(2)}$ kernel coefficients.   \\
 &&\\
   $ \mathcal{G}_{L_1,L_2,L_3}$& \ref{eq:CalG}& Coefficient from the integration of one 3-argument isotropic basis function. \\ 
  &&\\
  $ G^{L_1,L_2,L_3}_{M_1,M_2,M_3}$& \ref{eq:Gaunt_Coeff}& Gaunt coefficient from the integration of three spherical harmonic. \\ 
  &&\\
  $ \mathcal{H}_{L,L',L^"}^{\ell,m}$& \ref{eq:Fancy_H_def}& Coefficient from the integration of one 3-argument isotropic basis function times one spherical harmonic. \\ 
  &&\\
  $ \mathcal{J}_{L,L',L^"}^{\ell,\ell^",m,m^"}$&\ref{eq:Fancy_J_def}&Coefficient from the integration of one 3-argument isotropic basis function times two spherical harmonics.\\ 
  &&\\
  $ \overline{\mathcal{Q}}^{\Lambda,\Lambda_s,\Lambda'}$& $\S$\ref{sec:S_evaluation}& Coefficients from the evaluation of the $\hats$ integral for the five different configurations of the 1-loop covariance. \\   
  &&\\
 \hline
\end{tabular} 
\caption{Table with the coefficients relevant for the computation of the covariance. The first four coefficients have been taken from \cite{Ortola_4PCF}, although the fourth, $c_{j,n}^{(\mu)}$ has been modified for the work in this paper.}
\label{table:1}
\end{table}

\section{Modeling the 1-Loop Covariance with Second-Order Densities}\label{sec:Resulting_Cov}
Before presenting the detailed derivations, we outline the structure of the second-order contributions to the 4PCF covariance. The full set of terms can be grouped into two categories: fully-connected and partially-connected configurations. In total, five distinct diagrammatic structures arise. Three of them belong to the fully-connected class, shown in Figure~\ref{fig:dc_configurations}, while the remaining two fall into the partially-connected class, illustrated in Figure~\ref{fig:2nd_pc_terms}. 

The distinction between these categories is most easily seen in their diagrammatic representation. In the fully-connected case, Wick contractions link the two tetrahedra that define the covariance, ensuring that all density contrasts are connected across them. In contrast, in the partially-connected case, one loop is formed entirely within a single tetrahedron, leaving the connection between the two tetrahedra incomplete. We begin with the fully-connected configurations and then turn to the partially-connected ones.
\subsection{Fully-Connected Terms}
Within the class of fully-connected contributions, three distinct configurations arise, shown in Figure~\ref{fig:dc_configurations}. We evaluate them in the order displayed: the primary–secondary, half–half, and primary–primary contractions. These labels indicate how the second-order (“primary”) density contrasts are paired with the linear (“secondary”) densities through Wick’s theorem. In the diagrams, solid colored lines represent Wick contractions between densities of the same type, while dotted lines denote contractions between a primary and a secondary field. Each configuration corresponds to a different way of distributing the second-order density insertions across the two tetrahedra, and together they exhaust all possible fully-connected structures. We begin with the primary-secondary configuration. 

\subsubsection{Primary-Secondary Configuration}
Using Wick's theorem on Eq. (\ref{eq:sub_cov_before_Wicks}) for the primary-secondary (P-S) configuration shown in Fig. \ref{fig:dc_configurations} results in:
\begin{align}\label{eq:origin_of_twopi_to_15}
&{\rm Cov}_{4, \rm 1L}^{[2],{\rm P-S}}(\mathbf{R},\mathbf{R}') = \prod_{v=0}^{3} \int_{\vecs}\int_{\veck_v}\int_{\veck'_v}\left[ \;e^{-i\veck_v\cdot(\vecx+\vecr_v)} e^{-i\veck'_v \cdot(\vecx+\vecr'_v+\vecs)} \right]\nonumber \\
& \qquad \qquad \qquad \times\int d^3\vecq_1 \int d^3\vecq_2 \int d^3\vecq'_1 \int d^3\vecq'_2 \;\W(\vecq_1,\vecq_2) \nonumber \\
& \qquad \qquad \qquad \times  \W(\vecq'_1,\vecq'_2)\; \dD(\veck_0-\vecq_1-\vecq_2)\; \dD(\veck_0'-\vecq'_1-\vecq'_2)\nonumber \\
& \qquad \qquad \qquad \times  \left<\widetilde{\delta}_{\rm lin.}(\vecq_1)\widetilde{\delta}_{\rm lin.}(\veck'_3)\right>\left<\widetilde{\delta}_{\rm lin.}(\vecq_2)\widetilde{\delta}_{\rm lin.}(\veck'_2)\right> \left<\widetilde{\delta}_{\rm lin.}(\vecq'_1)\widetilde{\delta}_{\rm lin.}(\veck_3)\right>\nonumber \\
& \qquad \qquad \qquad \times\left<\widetilde{\delta}_{\rm lin.}(\vecq'_2)\widetilde{\delta}_{\rm lin.}(\veck_2)\right> \left<\widetilde{\delta}_{\rm lin.}(\veck_1)\widetilde{\delta}_{\rm lin.}(\veck'_1)\right> + 35\; {\rm perms.},
\end{align}
where $35$ perms. indicates the other 35 combinations\footnote{Finding the number of permutations is done in three steps. i) Choose one of the second-order densities in any of the two tetrahedra, then pair this second-order density with one of the linear densities from the other tetrahedron; one has 3 possibilities. ii) Repeat with the remaining second-order density; one has 2 possibilities. iii) Repeat with the opposite tetrahedron steps i) and ii); one has 6 possibilities. Total number of permutations = 3 $\times$ 2$\times$ 6  = 36 perms.} from Wick's theorem that we obtain for this primary-secondary structure. For brevity, we leave this out, but it is implicitly included into our calculations. Evaluating the ensemble averages as power spectra and performing the integrals $\vecq_s$ and $\vecq'_s$ results in:
\begin{align}
&{\rm Cov}_{4, \rm 1L}^{[2],{\rm P-S}}(\mathbf{R},\mathbf{R}') = (2\pi)^{15}\prod_{v=0}^{3}\int_{\vecs}\int_{\veck_v}\int_{\veck'_v}\left[ \;e^{-i\veck_v\cdot(\vecx+\vecr_v)} e^{-i\veck'_v \cdot(\vecx+\vecr'_v+\vecs)} \right]\nonumber \\
& \qquad \qquad \qquad \times \dD(\veck_0+\veck'_2+\veck'_3) \dD(\veck_0'+\veck_2+\veck_3) \dD(\veck_1 + \veck'_1) \nonumber \\
& \qquad \qquad \qquad \times \W(\veck'_3,\veck'_2) \W(\veck_3,\veck_2) \nonumber \\
& \qquad \qquad \qquad \times P_{\rm lin}(k_2) P_{\rm lin}(k_3) P_{\rm lin}(k'_2) P_{\rm lin}(k'_3) P_{\rm lin}(k_1),
\end{align}
where $(2\pi)^{15}$ arises from expressing the five ensemble average products of Eq. (\ref{eq:origin_of_twopi_to_15}) in terms of power spectra. By using the Dirac delta functions we obtain:
\begin{align}
&{\rm Cov}_{4, \rm 1L}^{[2],{\rm P-S}}(\mathbf{R},\mathbf{R}') = (2\pi)^{15} \int_{\vecs} \int_{\veck_1}e^{-i\veck_1\cdot(\vecr_1 - \vecs-\vecr'_1)} \Pk(k_1) \nonumber \\
& \qquad \qquad \qquad \times \int_{\veck_3}e^{-i\veck_3\cdot(\vecr_3 - \vecs-\vecr'_0)}P_{\rm lin}(k_3)\nonumber \\
& \qquad \qquad \qquad \times\int_{\veck'_3}e^{-i\veck'_3\cdot(\vecr'_3 + \vecs-\vecr_0)}P_{\rm lin}(k'_3)\nonumber \\
& \qquad \qquad \qquad \times \int_{\veck_2} \W(\veck_2,\veck_3)P_{\rm lin}(k_2)  e^{-i\veck_2\cdot(\vecr_2 - \vecs-\vecr'_0)}\nonumber \\
& \qquad \qquad \qquad \times \int_{\veck'_2}\W(\veck'_2,\veck'_3)P_{\rm lin}(k'_2)  e^{-i\veck'_2\cdot(\vecr'_2 + \vecs-\vecr_0)}.
\end{align}
To simplify the above result we will start by reducing the $\veck_2$ integral to have as low a dimension as possible :
\begin{equation}\label{eq:Def_of_I_2}
I^{(2)}(\mathbf{r}_2,-\mathbf{s},-\vecr'_0) \equiv \int_{\mathbf{k}_2} \W\left(\mathbf{k}_2, \mathbf{k}_3\right) P\left(k_2\right)  e^{-i \mathbf{k}_2 \cdot\left(\mathbf{r}_2-\mathbf{s}-\vecr'_0\right)}.
\end{equation}
We proceed by writing all the terms in the integrand in terms of the isotropic basis functions, which will allow us to separate the angular and radial parts; evaluating the rest of the $\mathbf{k}$ and $\mathbf{k}'$ integrals will follow the same procedure and their result is explicitly given in Eqs. (\ref{eq:I_PS_3})-(\ref{eq:I_PS_3'}). The second-order kernel $\W$ is already given in terms of the isotropic basis functions in Eq. (\ref{eq:W_2_def}). 

In the same manner, the complex exponential can be expanded into this basis using the plane wave expansion \cite{Ortola_4PCF}:
\begin{align}\label{eq:PWE}
&e^{-i \mathbf{k}_2 \cdot\left(\mathbf{r}_2-\mathbf{s}-\vecr'_0\right)} = (4\pi)^{3}\sum_{L_2,L_{2s},L'_{20}} C_{L_2,L_{2s},L'_{20}} \Upsilon_{L_2,L_{2s},L'_{20}} \; j_{L_2}(k_2 r_2)  j_{L_{2s}}(k_2 s)  \nonumber \\
& \qquad \qquad   \times j_{L'_{20}}(k_2 r'_0) \mathcal{P}_{L_2,L_{2s},L'_{20}}(\hatk_2,\hatk_2,\hatk_2) \mathcal{P}_{L_2,L_{2s},L'_{20}}(\hatr_2,-\hats,-\hatr'_0).  
\end{align}
We define the PWE constants as:
\begin{align} \label{eq:Constant_from_exp}
C_{\ell'_1,\ell'_2,\ell'_3} \equiv i^{\ell'_1+\ell'_2+\ell'_3} \; \sqrt{(2\ell'_1+1)(2\ell'_2+1)(2\ell'_3+1)}, 
 \end{align}
and
\begin{align}\label{eq:Upsilon_def}
 \Upsilon_{L_1,L_2,L_3} =  \quad \frac{(-1)^{L_1+L_2+L_3}}{\sqrt{(2L_1+1)(2L_2+1)(2L_3+1)}}. 
\end{align}
Finally, with all of the terms in the isotropic basis, after evaluating the angular integrals we obtain:
\begin{align}\label{eq:I_PS_2}
&I_{j,n}^{(2)}(\mathbf{r}_2,-\mathbf{s},-\vecr'_0) = (4\pi)^{2} \sum_{L_2,L_{2s},L'_{20}} C_{L_2,L_{2s},L'_{20}} \Upsilon_{L_2,L_{2s},L'_{20}} \mathcal{H}_{L_2,L_{2s},L'_{20}}^{j,m_j}  \nonumber \\
& \qquad \qquad \qquad \qquad \qquad  \times g^{[n]}_{L_2,L_{2s},L'_{20}}(r_2,s,r'_0) \mathcal{P}_{L_2,L_{2s},L'_{20}}(\hatr_2,-\hats,-\hatr'_0),
\end{align}
where we have intentionally left out the $j$, $n$ sums and factor of $4\pi$ since they will also affect the result of the $\veck_3$ integral; to denote this, we have included the subscripts $j$, $n$ on the left-hand side of the above equation. We reintroduce these sums in Eq. (\ref{eq:PS_res}). The constant $\mathcal{H}$ comes directly from evaluating the angular integral as shown in Eq. (\ref{eq:Fancy_H_def}). The radial integral $g$ is defined as:
\begin{align}\label{eq:gint}
g_{L,L',L''}^{[n]}(r, r',r'') \equiv \int \frac{dk}{2\pi^2} \;k^{n+2} j_{L}(k r) j_{L'}(k r') j_{L''}(k r'') \;\Pk(k).
\end{align}
\begin{figure}[h!]
\centering
\includegraphics[scale=0.8]{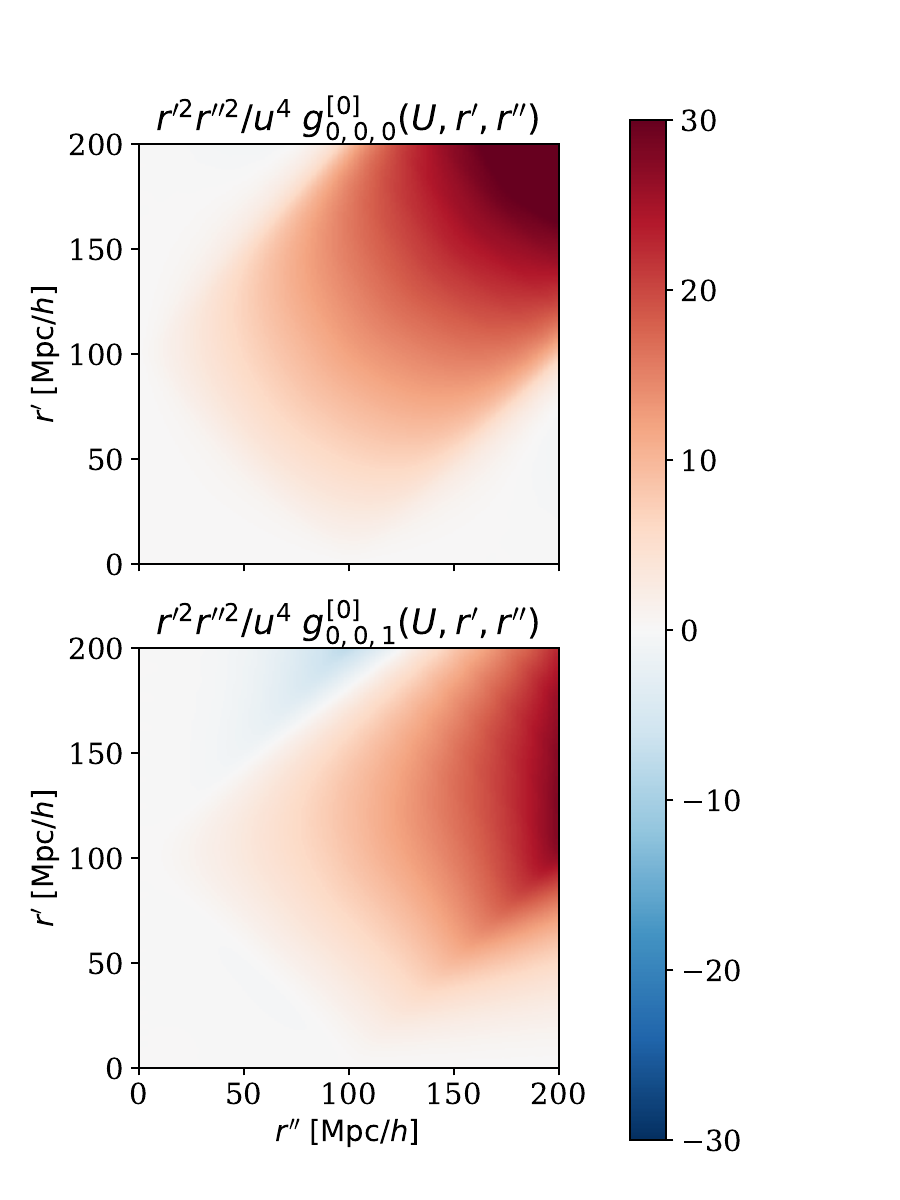}
\caption{Here, we show Eq. (\ref{eq:gint}) for fixed $r' = U \equiv 100 \;\left[{\rm Mpc}/h\right]$, $n' = 0$ $\ell'=0$,  $\ell = 0$ and $L = \left\{0,1\right\}$. The plot has been taken from section 3 of \cite{Ortola_4PCF}. The \textit{upper panel} shows the integral for $L=0$, while the \textit{lower panel} shows the integral for $L=1$. Since the 4PCF can be approximated as the square of the 2PCF on large scales, $(\xi_{0})^{2}(r) \sim (1/r^2)^2$, we have weighted the integral by $r^2r_i^2/u^{4}$, with $u \equiv 10 \;\left[{\rm Mpc}/h\right]$ to take out its fall-off. Both the \textit{upper} and \textit{lower} panel show the behavior of the integral creates a rectangular boundary. Analytical results with the power spectrum using a power law and explaining the rectangular boundaries ($r' = r^"+U$, $r' = -r^"+U$, and $r' = r^"-U$) have been obtained in \cite{Ortola_4PCF}; further results are also evaluated in \cite{Chellino_3SBF}.}
\label{fig:gint2}
\end{figure}

Eq. (\ref{eq:I_PS_2}) immediately allows us to evaluate the $\veck_3,\;\veck'_2$ and $\veck'_3$ integrals as:
\begin{align}\label{eq:I_PS_3}
&I_{j,n}^{(3)}(\mathbf{r}_3,-\mathbf{s},-\vecr'_0) = (4\pi)^{2} \sum_{L_3,L_{3s},L'_{30}} C_{L_3,L_{3s},L'_{30}} \Upsilon_{L_3,L_{3s},L'_{30}} \mathcal{H}_{L_3,L_{3s},L'_{30}}^{j,m_j}  \nonumber \\
& \qquad \qquad \qquad \qquad \qquad  \times g^{[-n]}_{L_3,L_{3s},L'_{30}}(r_3,s,r'_0) \mathcal{P}_{L_3,L_{3s},L'_{30}}(\hatr_3,-\hats,-\hatr'_0),
\end{align}
\begin{align}\label{eq:I_PS_2'}
&I_{j',n'}^{(2')}(\vecr_0,-\mathbf{s},-\vecr'_2) = (4\pi)^{2} \sum_{L_{20},L'_{2s},L'_2} C_{L_{20},L'_{2s},L'_2} \Upsilon_{L_{20},L'_{2s},L'_2} \mathcal{H}_{L_{20},L'_{2s},L'_2}^{j',m_{j'}}  \nonumber \\
& \qquad \qquad \qquad \qquad \qquad  \times g^{[n']}_{L_{20},L'_{2s},L'_2}(r_0,s,r'_2) \mathcal{P}_{L_{20},L'_{2s},L'_2}(\hatr_0,-\hats,-\hatr'_2),
\end{align}
\begin{align}\label{eq:I_PS_3'}
&I_{j',n'}^{(3')}(\vecr_0,-\vecs,-\vecr'_3) = (4\pi)^{2} \sum_{L_0,L'_{3s},L'_3} C_{L_0,L'_{3s},L'_3} \Upsilon_{L_0,L'_{3s},L'_3} \mathcal{H}_{L_0,L'_{3s},L'_3}^{j',m_{j'}}  \nonumber \\
& \qquad \qquad \qquad \qquad \qquad  \times g^{[-n']}_{L_0,L'_{3s},L'_3}(r_0,s,r'_3) \mathcal{P}_{L_0,L'_{3s},L'_3}(\hatr_0,-\hats,-\hatr'_3).
\end{align}
  
Next, we must evaluate the $\veck_1$ integral. To do so, we follow the same procedure as before but without the second-order kernel. This means that the evaluation of its angular integral will produce a modified Gaunt coefficient instead of an $\mathcal{H}$ coefficient. We find:
\begin{align}\label{eq:I_PS_1}
&I^{(1)}(\vecr_1,-\vecr'_1,-\vecs) =(4\pi)^2 \sum_{L_1,L_{1s},L'_{1}} C_{L_1,L_{1s},L'_{1}} \Upsilon_{L_1,L_{1s},L'_{1}} \mathcal{G}_{L_1,L_{1s},L'_{1}}  \nonumber \\
& \qquad \qquad \qquad \qquad \qquad  \times g^{[0]}_{L_1,L_{1s},L'_{1}}(r_1,s,r'_1) \mathcal{P}_{L_1,L_{1s},L'_{1}}(\hatr_1,-\hats,-\hatr'_1),
\end{align}
with $\mathcal{G}$ defined in Eq. (\ref{eq:CalG}). Finally, after evaluating the $\hats$ integral using Eq. (\ref{eq:s_hat_PS}), we obtain:
\begin{empheq}[box=\widefbox]{align}\label{eq:PS_res}
&{\rm Cov}_{4, \rm 1L}^{[2],{\rm P-S}} (\mathbf{R},\mathbf{R}') = (2\pi)^{15} (4\pi)^{12} \sum_{\rm All}  \nonumber \\
& \qquad \qquad \times c_{j,n}^{(\mu)} c_{j',n'}^{(\beta)}\;\overline{\mathcal{Q}}^{\Lambda,\Lambda_s,\Lambda'}_{\rm (P-S)} \nonumber \\
& \qquad \qquad \times C_{L_1,L'_1,L_{1s}} \Upsilon_{L_1,L'_1,L_{1s}} \mathcal{G}_{L_1,L'_1,L_{1s}} \nonumber \\
& \qquad \qquad \times C_{L_2,L_{2s},L'_{20}} \Upsilon_{L_2,L_{2s},L'_{20}} \mathcal{H}_{L_2,L_{2s},L'_{20}}^{j,m_j}\nonumber \\
& \qquad \qquad \times  C_{L_3,L_{3s},L'_{30}} \Upsilon_{L_3,L_{3s},L'_{30}}\mathcal{H}_{L_3,L_{3s},L'_{30}}^{j,m_j} \nonumber \\
& \qquad \qquad \times C_{L'_2,L'_{2s},L_{20}} \Upsilon_{L'_2,L'_{2s},L_{20}} \mathcal{H}_{L'_2,L'_{2s},L_{20}}^{j',m_{j'}} \nonumber \\
& \qquad \qquad \times C_{L'_3,L'_{3s},L_{30}} \Upsilon_{L'_3,L'_{3s},L_{30}} \mathcal{H}_{L'_3,L'_{3s},L_{30}}^{j',m_{j'}}\nonumber \\
& \qquad \qquad \times S_{\{L\}, {\rm P-S}}^{(\{L_{is}\})}(r_0,r'_0,r_1,r'_1,r_2,r'_2,r_3,r'_3) \nonumber \\
& \qquad \qquad \times \PP_{\Lambda}(\hatR)\PP_{\Lambda'}(\hatR') +\;35\;{\rm perms.},
\end{empheq}
where the $\Lambda,\; \Lambda_s$ and $\Lambda'$ momenta have been defined in $\S$\ref{sec:s_hat_PS_Evaluation}. Table \ref{table:1} provides the equations where each constant is defined. We have defined the radial integral $S$ as:
\begin{align}\label{eq:Sint_PS}
&S_{\{L\}, {\rm P-S}}^{(\{L_{is}\})}(r_0,r'_0,r_1,r'_1,r_2,r'_2,r_3,r'_3) \nonumber \\
& \qquad  \equiv \int ds\;s^2\; g_{L_1,L'_1,L1s}^{[0]} (r_1,r'_1,s) g_{L_2,L_{2s},L'_{20}}^{[n]} (r_2,s,r'_{0})\nonumber \\
& \qquad  \times g_{L_3,L_{3s},L'_{30}}^{[-n]} (r_3,s,r'_{0}) 
g_{L'_2,L'_{2s},L_{20}}^{[n']} (r'_2,s,r_{0})g_{L'_3,L'_{3s},L_{30}}^{[-n']} (r'_3,s,r_{0}).
\end{align}
The subscript $\{L\}$ denotes the set of all angular momentum indices that affect one of the resulting variables (\textit{i.e.}, the $r_0$, $r'_0$,$\cdots$ variables). The superscript $(\{L_{is}\})$ denotes the set of all angular momentum indices that are coupled to $s$ in the spherical Bessel functions. 

\begin{figure}[h!]
\centering
\includegraphics[scale=0.7]{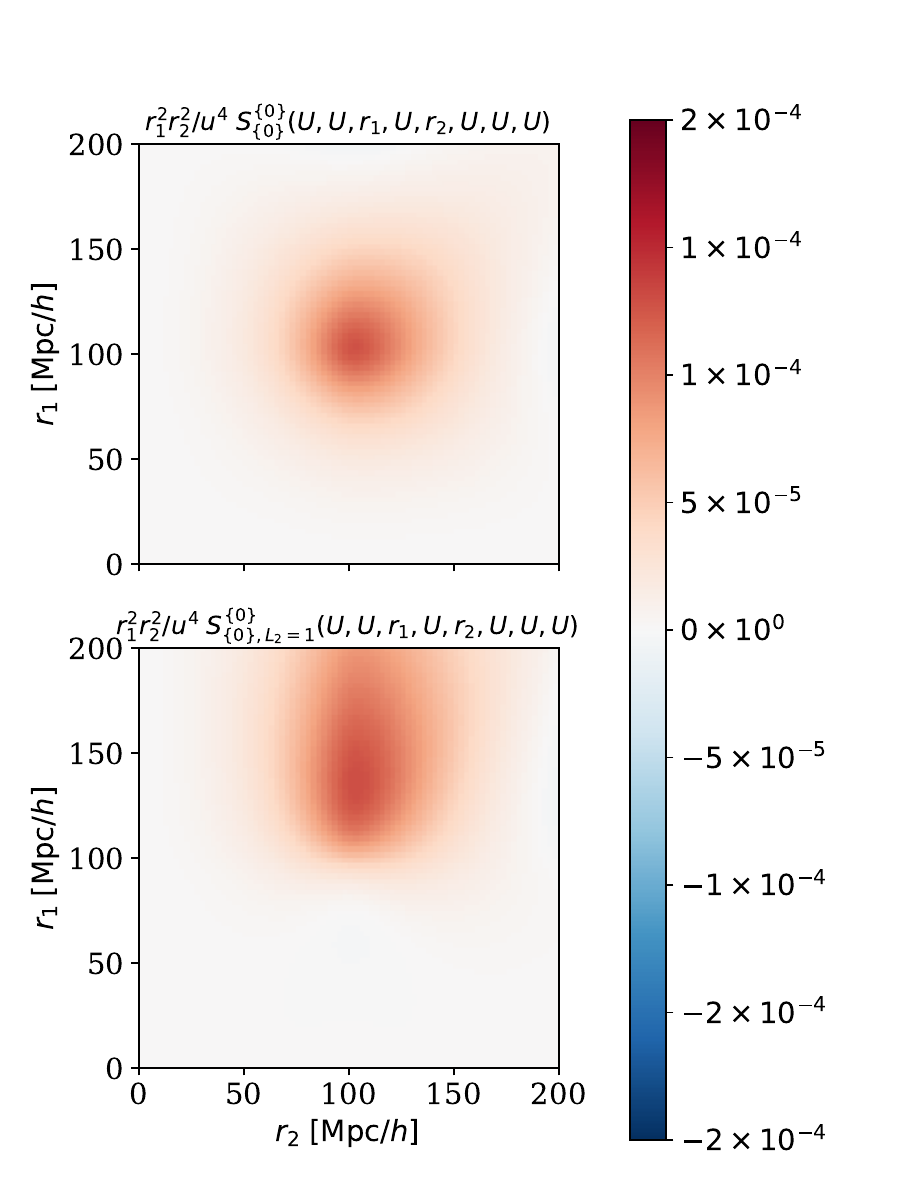}
\caption{Here, we show Eqs. (\ref{eq:Sint_PS}), (\ref{eq:Sint_HH}) and (\ref{eq:Sint_PP}) with fixed variables at $ U \equiv 100 \;\left[{\rm Mpc}/h\right]$, except for $r_1$ and $r_2$. We also chose to evaluate $n' = n=0$ and all angular momentum $\{L\} = 0$, except for $L_2 = \left\{0,1\right\}$. The \textit{upper panel} shows the integral for $L_2=0$, while the \textit{lower panel} shows the integral for $L_2=1$. Since the 4PCF can be approximated as the square of the 2PCF on large scales, $(\xi_{0})^{2}(r) \sim (1/r^2)^2$, we have weighted the integral by $r^2r_i^2/u^{4}$, with $u \equiv 10 \;\left[{\rm Mpc}/h\right]$, to take out its fall-off. The $S$ integral is composed of the intermediate radial integrals, $g$, which reveals a rectangular boundary as shown in Fig. \ref{fig:gint2}. The integration of the different rectangular boundaries creates the oval-shaped regions shown in both panels. Analytical results for similar radial integrals with the power spectrum being a power law have been obtained in \cite{Ortola_4PCF}.}
\label{fig:Sint}
\end{figure}

\subsubsection{Half-Half Configuration}
Next, we analyze the half–half configuration. The name reflects the fact that its Wick contractions split evenly: one contraction joins two primary densities, while another pairs a primary with a secondary density. Using Wick's theorem on Eq. (\ref{eq:sub_cov_before_Wicks}), which describes the half-half configuration shown in Fig. \ref{fig:dc_configurations}, we find:
\begin{align}
&{\rm Cov}_{4, \rm 1L}^{[2],{\rm H-H}}(\mathbf{R},\mathbf{R}') = \prod_{v=0}^{3} \int_{\vecs}\int_{\veck_v}\int_{\veck'_v}\left[\;e^{-i\veck_v\cdot(\vecx+\vecr_v)} e^{-i\veck'_v \cdot(\vecx+\vecr'_v+\vecs)}\right]\nonumber \\
& \qquad \qquad \qquad \times  \int d^3\vecq_1 \int d^3\vecq_2 \int d^3\vecq'_1 \int d^3\vecq'_2 \; \W(\vecq_1,\vecq_2) \nonumber \\
& \qquad \qquad \qquad \times \W(\vecq'_1,\vecq'_2)\; \dD(\veck_0-\vecq_1-\vecq_2)\; \dD(\veck_0'-\vecq'_1-\vecq'_2)\nonumber \\
& \qquad \qquad \qquad \times  \left<\widetilde{\delta}_{\rm lin.}(\vecq_1)\widetilde{\delta}_{\rm lin.}(\veck'_3)\right>\left<\widetilde{\delta}_{\rm lin.}(\vecq_2)\widetilde{\delta}_{\rm lin.}(\vecq'_1)\right> \left<\widetilde{\delta}_{\rm lin.}(\vecq'_2)\widetilde{\delta}_{\rm lin.}(\veck_3)\right>\nonumber \\
& \qquad \qquad \qquad \times\left<\widetilde{\delta}_{\rm lin.}(\veck_1)\widetilde{\delta}_{\rm lin.}(\veck'_1)\right> \left<\widetilde{\delta}_{\rm lin.}(\veck_2)\widetilde{\delta}_{\rm lin.}(\veck'_2)\right> + 89{\rm\; perms.},
\end{align}
where $89$ perms. indicates the other combinations\footnote{Finding the number of permutations is done in four steps. i) Choose one of the second-order densities in either of the two tetrahedra, then pair this second-order density with any densities from the other tetrahedron; one has 5 possibilities. ii) Pair the remaining second-order density with any of the linear densities in the other tetrahedron; one has 3 possibilities. iii) In the other tetrahedron, pair the remaining second-order density with any of the linear densities in the first tetrahedron; one has 3 possibilities. iv) Choose one of the linear densities in any of the two tetrahedra, then pair this density with a linear density in the other tetrahedron; one has 2 possibilities. Total number of permutations = 5 $\times$ 3$\times$ 3 $\times$ 2  = 90 perms.} due to Wick's theorem that we can obtain for this primary-secondary structure. For brevity, we leave this out, but it is implicitly included into our calculations. Since the $\veck_1$, $\veck_2$, $\veck'_1$, and $\veck'_2$ integrals do not depend on any second-order kernel, evaluating them will give the same result as in Eq. (\ref{eq:I_PS_1}). 

Then, performing the $\vecq_1$, $\vecq_2$, $\vecq'_2$, $\veck_0$ and $\veck_0'$ integrals results in:
\begin{align}
&{\rm Cov}_{4, \rm 1L}^{[2],{\rm H-H}}(\mathbf{R},\mathbf{R}') = (2\pi)^{15}\;\int_{\vecs} I^{(1)}(\vecr_1, -\vecs,- \vecr'_1)I^{(2)}(\vecr_2,-\vecs, - \vecr'_2)\nonumber \\
& \qquad \qquad \qquad \times \int_{\veck_3}\int_{\veck'_3}\int_{\vecq'_1}\left[ e^{-i\veck_3\cdot (\vecr_3-\vecs-\vecr'_0) } e^{-i\veck'_3\cdot (\vecr'_3+\vecs-\vecr_0)} e^{-i\vecq'_1\cdot(\vecr'_0-\vecr_0+\vecs)} \right]\nonumber \\
& \qquad \qquad \qquad \times  \W(\vecq'_1,-\veck_3)  \W(\vecq'_1, \veck'_3 ) P_{\rm lin}(q'_1)P_{\rm lin}(k'_3)P_{\rm lin}(k_3).
\end{align}
Expressing the second-order kernels in the isotropic basis as in Eq.(\ref{eq:W_2_def}), and performing the same analysis as in Eqs. (\ref{eq:Def_of_I_2})-(\ref{eq:I_PS_2}) for the $\veck_3$, $\veck'_3$ and $\vecq'_1$ integrals, we obtain:
\begin{align}\label{eq:I_PM_3}
&I_{j,n}^{(3)}(\mathbf{r}_3,-\mathbf{s},-\vecr'_0) = (4\pi)^{2} \sum_{L_3,L_{3s},L'_{30}} C_{L_3,L_{3s},L'_{30}} \Upsilon_{L_3,L_{3s},L'_{30}} \mathcal{H}_{L_3,L_{3s},L'_{30}}^{j,m_j}  \nonumber \\
& \qquad \qquad \qquad \qquad \qquad  \times g^{[-n]}_{L_3,L_{3s},L'_{30}}(r_3,s,r'_0) \mathcal{P}_{L_3,L_{3s},L'_{30}}(\hatr_3,-\hats,-\hatr'_0),
\end{align}
\begin{align}\label{eq:I_PM_3'}
&I_{j',n'}^{(3')}(\mathbf{r}'_3,\mathbf{s},-\vecr_0) = (4\pi)^{2} \sum_{L'_3,L'_{3s},L_{30}} C_{L'_3,L'_{3s},L_{30}} \Upsilon_{L'_3,L'_{3s},L_{30}} \mathcal{H}_{L'_3,L'_{3s},L_{30}}^{j',m_{j'}}  \nonumber \\
& \qquad \qquad \qquad \qquad \qquad  \times g^{[-n']}_{L'_3,L'_{3s},L_{30}}(r'_3,s,r_0) \mathcal{P}_{L'_3,L'_{3s},L_{30}}(\hatr'_3,\hats,-\hatr_0),
\end{align}
\begin{align}\label{eq:I_PM_q1'}
&I_{j',n'}^{(q'_1)}(\mathbf{r}'_0,\mathbf{s},-\vecr_0) = (4\pi)^{2} \sum_{L'_{q'_0},L'_{qs},L_{q0}} C_{L'_{q'_0},L'_{qs},L_{q0}} \Upsilon_{L'_{q'_0},L'_{qs},L_{q0}} \mathcal{J}_{L'_{q'_0},L'_{qs},L_{q0}}^{j,j',m_j,m_{j'}}  \nonumber \\
& \qquad \qquad \qquad \qquad \qquad  \times g^{[n+n']}_{L'_{q'_0},L'_{qs},L_{q0}}(r'_0,s,r_0) \mathcal{P}_{L'_{q'_0},L'_{qs},L_{q0}}(\hatr'_0,\hats,-\hatr_0),
\end{align}
where evaluating the angular integral for $\vecq'_1$ resulted in the constant $\mathcal{J}$. This integral involved the evaluation of one 3-argument isotropic basis function and two spherical harmonics, as shown in Eq. (\ref{eq:Fancy_J_def}). Therefore, the half-half configuration results in:
\begin{empheq}[box=\widefbox]{align}\label{eq:HH_res}
&{\rm Cov}_{4, \rm 1L}^{[2],{\rm H-H}} (\mathbf{R},\mathbf{R}') = (2\pi)^{15} (4\pi)^{12} \sum_{\rm All}  \nonumber \\
& \qquad \qquad \times c_{j,n}^{(\mu)} c_{j',n'}^{(\beta)}\overline{\mathcal{Q}}^{\Lambda,\Lambda_s,\Lambda'}_{\rm (H-H)}\nonumber \\
& \qquad \qquad \times C_{L_1,L'_1,L_{1s}} \Upsilon_{L_1,L'_1,L_{1s}} \mathcal{G}_{L_1,L'_1,L_{1s}} \nonumber \\
& \qquad \qquad \times C_{L_2,L'_2,L_{2s}} \Upsilon_{L_2,L'_2,L_{2s}} \mathcal{G}_{L_2,L'_2,L_{2s}}\nonumber \\
& \qquad \qquad \times  C_{L_3,L_{3s},L'_{30}} \Upsilon_{L_3,L_{3s},L'_{30}} \mathcal{H}_{L_3,L_{3s},L'_{30}}^{j,m_j} \nonumber \\
& \qquad \qquad \times C_{L'_3,L'_{3s},L_{30}} \Upsilon_{L'_3,L'_{3s},L_{30}} \mathcal{H}_{L'_3,L'_{3s},L_{30}}^{j',m_{j'}}\nonumber \\
& \qquad \qquad \times C_{L'_{q'_0},L'_{qs},L_{q0}} \Upsilon_{L'_{q'_0},L'_{qs},L_{q0}} \mathcal{J}_{L'_{q'_0},L'_{qs},L_{q0}}^{j,j',m_j,m_{j'}}\nonumber \\
& \qquad \qquad \times S_{\{L\}, {\rm H-H}}^{(\{L_{is}\})}(r_0,r'_0,r_1,r'_1,r_2,r'_2,r_3,r'_3) \nonumber \\
& \qquad \qquad \times\PP_{\Lambda}(\hatR)\PP_{\Lambda'}(\hatR')+\;89\;{\rm perms.},
\end{empheq}
where the $\Lambda,\; \Lambda_s$ and $\Lambda'$ momenta have been defined in $\S$\ref{sec:s_hat_HH_Evaluation}. Table \ref{table:1} provides the equation where all the constants have been defined. We have defined the radial integral $S$ as:
\begin{align}\label{eq:Sint_HH}
&S_{\{L\}, {\rm H-H}}^{(\{L_{is}\})}(r_0,r'_0,r_1,r'_1,r_2,r'_2,r_3,r'_3) \nonumber \\
& \qquad  \equiv \int ds\;s^2\; g_{L_1,L'_1,L1s}^{[0]} (r_1,r'_1,s) g_{L_2,L'_2,L2s}^{[0]} (r_2,r'_2,s)\nonumber \\
& \qquad  \times g_{L_3,L_{3s},L'_{30}}^{[-n]} (r_3,s,r'_{0}) g_{L'_3,L'_{3s},L_{30}}^{[-n']} (r'_3,s,r_{0}) g_{L'_{q'_0},L'_{qs},L_{q0}}^{[n+n']} (r'_0,s,r_{0}).
\end{align} 
The subscript $\{L\}$ denotes the set of all angular momentum indices that affect one of the resulting variables (\textit{i.e.}, the $r_0$, $r'_0$,$\cdots$ variables). The superscript $(\{L_{is}\})$ denotes the set of all angular momentum indices that are coupled to $s$ in the spherical Bessel functions. 

\subsubsection{Primary-Primary Configuration}

Using Wick's theorem on Eq. (\ref{eq:sub_cov_before_Wicks}), which describes the Primary-Primary configuration shown in Fig. \ref{fig:dc_configurations}, we find:
\begin{align}
&{\rm Cov}_{4, \rm 1L}^{[2],{\rm P-P}}(\mathbf{R},\mathbf{R}') = \prod_{v=0}^{3} \int_{\vecs}\int_{\veck_v}\int_{\veck'_j=v}\left[\;e^{-i\veck_v\cdot(\vecx+\vecr_v)} e^{-i\veck'_v \cdot(\vecx+\vecr'_v+\vecs)}\right]\nonumber \\
& \qquad \qquad \qquad \times \int d^3\vecq_1 \int d^3\vecq_2 \int d^3\vecq'_1 \int d^3\vecq'_2 \; \W(\vecq_1,\vecq_2) \nonumber \\
& \qquad \qquad \qquad \times \W(\vecq'_1,\vecq'_2)\; \dD(\veck_0-\vecq_1-\vecq_2)\; \dD(\veck_0'-\vecq'_1-\vecq'_2)\nonumber \\
& \qquad \qquad \qquad \times  \left<\widetilde{\delta}_{\rm lin.}(\vecq_1)\widetilde{\delta}_{\rm lin.}(\vecq'_1)\right>\left<\widetilde{\delta}_{\rm lin.}(\vecq_2)\widetilde{\delta}_{\rm lin.}(\vecq'_2)\right> \left<\widetilde{\delta}_{\rm lin.}(\veck_1)\widetilde{\delta}_{\rm lin.}(\veck'_1)\right>\nonumber \\
& \qquad \qquad \qquad \times\left<\widetilde{\delta}_{\rm lin.}(\veck_2)\widetilde{\delta}_{\rm lin.}(\veck'_2)\right> \left<\widetilde{\delta}_{\rm lin.}(\veck_3)\widetilde{\delta}_{\rm lin.}(\veck'_3)\right> + 11{\rm \; perms.},
\end{align}
where $11$ perms. indicates the other 11 combinations\footnote{Finding the number of permutations is done in three steps. i) Choose one of the second-order densities in either of the two tetrahedra, then pair this second-order density with another second-order density from the other tetrahedron; one has 2 possibilities. ii) Pair the remaining second-order density with the other second-order density in the other tetrahedron; one has 1 possibilities. iii) Pair the remaining linear densities with a linear density in the other tetrahedron; one has 6 possibilities. Total number of permutations = 2 $\times$ 1$\times$ 6 = 12 perms.}of Wick's theorem that we can obtain for this primary-primary structure. For brevity, we leave this out, but it is implicitly included into our calculations. Since the $\veck_1$, $\veck_2$, $\veck_3$ $\veck'_1$, $\veck'_2$, and $\veck'_3$ integrals do not contain any second-order kernel, evaluating them will give the same result as in Eq. (\ref{eq:I_PS_1}). 

Then, performing the $\vecq'_1$, $\vecq'_2$, $\veck'_0$, and $\veck_0$ integrals yields:
\begin{align}\label{eq:PP_uncopling}
&{\rm Cov}_{4, \rm 1L}^{[2],{\rm P-P}}(\mathbf{R},\mathbf{R}') = (2\pi)^{15}\;\int_{\vecs} I^{(1)}(\vecr_1, - \vecr'_1, - \vecs)I^{(2)}(\vecr_2, - \vecr'_2, - \vecs)\nonumber \\
& \qquad \qquad \qquad \times I^{(3)}(\vecr_3, - \vecr'_3, - \vecs)\int_{\vecq_1}\int_{\vecq_2}\left[ e^{-i\vecq_1\cdot (\vecr_0-\vecr'_0-\vecs) } e^{-i\vecq_2\cdot (\vecr_0-\vecr'_0-\vecs)} \right]\nonumber \\
& \qquad \qquad \qquad \times  \W(\vecq_1,\vecq_2)  \W(\vecq_1, \vecq_2 )  P_{\rm lin}(q_1)P_{\rm lin}(q_2).
\end{align}
Expressing the two second-order kernels in terms of the isotropic basis function as in Eq. (\ref{eq:W_2_def}), and performing the same analysis as in Eqs. (\ref{eq:Def_of_I_2})-(\ref{eq:I_PS_2}) for the $\vecq_1$, and $\vecq_2$ integrals, we obtain:
\begin{align}\label{eq:I_PP_q1}
&I_{j,j',n,n'}^{(q_1)}(\vecr_0,-\vecr'_0,-\vecs) = (4\pi)^{2} \sum_{L_{01},L'_{01},L'_{1qs}} C_{L_{01},L'_{01},L'_{1qs}} \Upsilon_{L_{01},L'_{01},L'_{1qs}} \mathcal{J}_{L_{01},L'_{01},L'_{1qs}}^{j,j',m_j,m_{j'}}  \nonumber \\
& \qquad \qquad \qquad \qquad \quad  \times g^{[n+n']}_{L_{01},L'_{01},L'_{1qs}}(r_0,r'_0,s) \mathcal{P}_{L_{01},L'_{01},L'_{1qs}}(\hatr_0,-\hatr'_0,-\hats),
\end{align}
and
\begin{align}\label{eq:I_PP_q2}
&I_{j,j',n,n'}^{(q_2)}(\vecr_0,-\vecr'_0,-\vecs) = (4\pi)^{2} \sum_{L_{02},L'_{02},L'_{2qs}} C_{L_{02},L'_{02},L'_{2qs}} \Upsilon_{L_{02},L'_{02},L'_{2qs}} \mathcal{J}_{L_{02},L'_{02},L'_{2qs}}^{j,j',m_j,m_{j'}}  \nonumber \\
& \qquad \qquad \qquad \qquad \quad  \times g^{[-n-n']}_{L_{02},L'_{02},L'_{2qs}}(r_0,r'_0,s) \mathcal{P}_{L_{02},L'_{02},L'_{2qs}}(\hatr_0,-\hatr'_0,-\hats),
\end{align}
where the evaluation of the angular integrals $\vecq'_1$ and $\vecq'_2$ resulted in the constant $\mathcal{J}$. This integral involved the evaluation of one 3-argument isotropic basis function and two spherical harmonics, as shown in Eq. (\ref{eq:Fancy_J_def}). Therefore, the primary-primary configuration, results in:
\begin{empheq}[box=\widefbox]{align}\label{eq:PP_result}
&{\rm Cov}_{4, \rm 1L}^{[2],{\rm P-P}} (\mathbf{R},\mathbf{R}') = (2\pi)^{15} (4\pi)^{12} \sum_{\rm All}\nonumber \\
& \qquad \qquad \times c_{j,n}^{(\mu)}c_{j',n'}^{(\beta)}\overline{\mathcal{Q}}^{\Lambda,\Lambda_s,\Lambda'}_{\rm (P-P)}  \nonumber \\
& \qquad \qquad \times C_{L_1,L'_1,L_{1s}} \Upsilon_{L_1,L'_1,L_{1s}} \mathcal{G}_{L_1,L'_1,L_{1s}} \nonumber \\
& \qquad \qquad \times C_{L_2,L'_2,L_{2s}} \Upsilon_{L_2,L'_2,L_{2s}} \mathcal{G}_{L_2,L'_2,L_{2s}}\nonumber \\
& \qquad \qquad \times  C_{L_3,L'_{3},L_{3s}} \Upsilon_{L_3,L'_{3},L_{3s}} \mathcal{G}_{L_3,L'_{3},L_{3s}} \nonumber \\
& \qquad \qquad \times C_{L_{01},L'_{01},L'_{1qs}} \Upsilon_{L_{01},L'_{01},L'_{1qs}} \mathcal{J}_{L_{01},L'_{01},L'_{1qs}}^{j,j',m_j,m_{j'}} \nonumber \\
& \qquad \qquad \times C_{L_{02},L'_{02},L'_{2qs}} \Upsilon_{L_{02},L'_{02},L'_{2qs}} \mathcal{J}_{L_{02},L'_{02},L'_{2qs}}^{j,j',m_j,m_{j'}}\nonumber \\
& \qquad \qquad \times S_{\{L\}, {\rm P-P}}^{(\{L_{is}\})}(r_0,r'_0,r_1,r'_1,r_2,r'_2,r_3,r'_3) \nonumber \\
& \qquad \qquad \times\PP_{\Lambda}(\hatR)\PP_{\Lambda'}(\hatR')+\;11\;{\rm perms.},
\end{empheq}
where the $\Lambda,\; \Lambda_s$ and $\Lambda'$ momenta have been defined in $\S$\ref{sec:s_hat_PP_Evaluation}. Table \ref{table:1} provides the equation where all the constants have been defined. We have defined the radial integral $S$ as:
\begin{align}\label{eq:Sint_PP}
&S_{\{L\}, {\rm P-P}}^{(\{L_{is}\})}(r_0,r'_0,r_1,r'_1,r_2,r'_2,r_3,r'_3) \nonumber \\
& \qquad  \equiv \int ds\;s^2\; g_{L_1,L'_1,L1s}^{[0]} (r_1,r'_1,s) g_{L_2,L'_2,L2s}^{[0]} (r_2,r'_2,s)\nonumber \\
& \qquad  \times g_{L_3,L'_{3},L_{3s}}^{[0]} (r_3,r'_3,s) g_{L_{01},L'_{01},L_{1qs}}^{[n+n']} (r_0,r'_0,s) g_{L_{02},L'_{02},L_{2qs}}^{[-n-n']} (r_0,r'_{0},s).
\end{align} 
The subscript $\{L\}$ denotes the set of all angular momentum indices that affect one of the resulting variables (\textit{i.e.}, the $r_0$, $r'_0$,$\cdots$ variables). The superscript $(\{L_{is}\})$ denotes the set of all angular momentum indices that are coupled to $s$ in the spherical Bessel functions. 
\subsection{Partially-Connected Terms}
\begin{figure}[h!]
\centering
\includegraphics[scale=0.2]{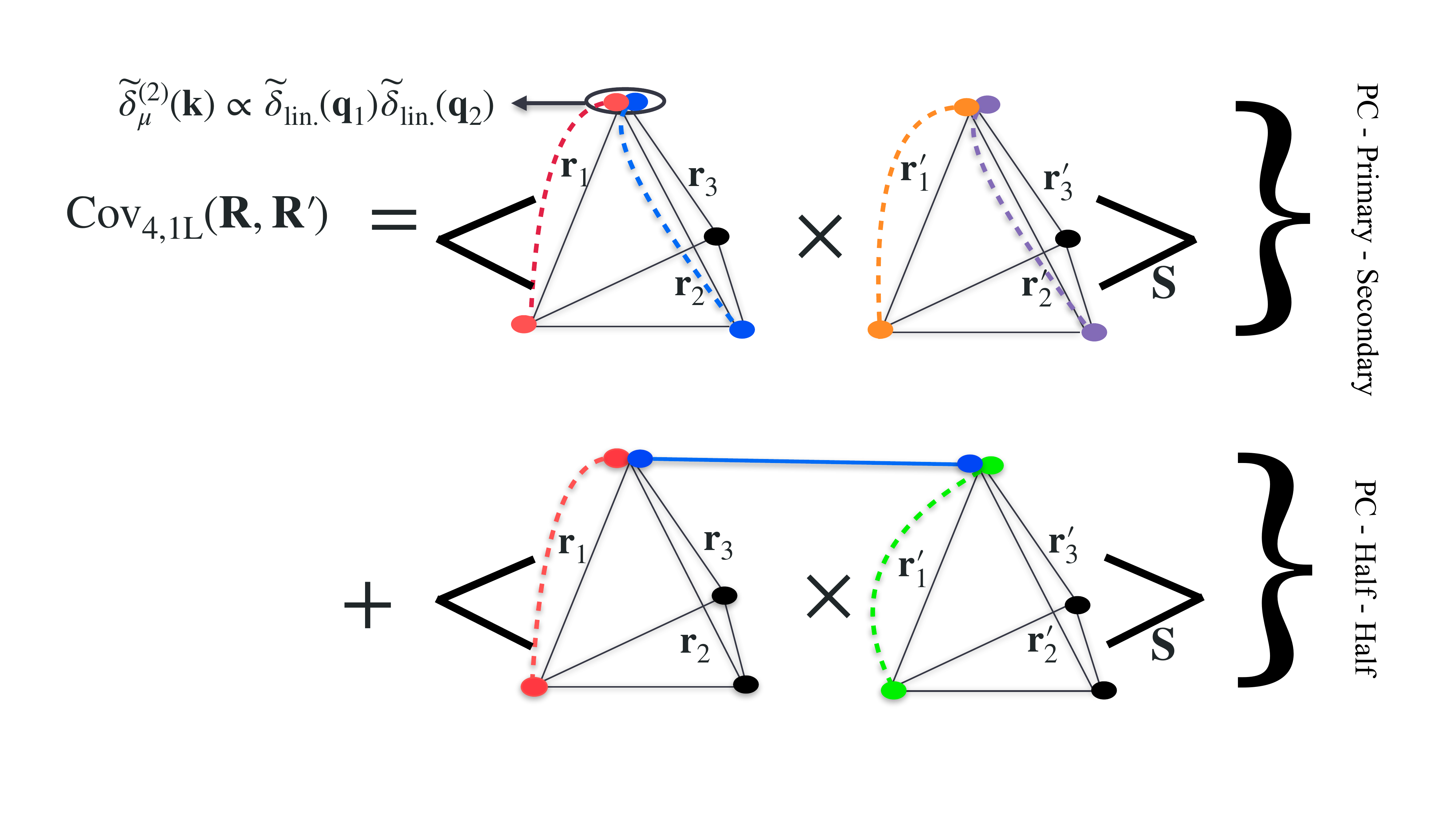}
\caption{Illustration of the two different possible configurations of the partially-connected covariance at 1-loop order. They are: PC-Primary-Secondary and PC-Half-Half. The solid lines represent the contraction of two primary density contrasts via Wick's theorem; the dashed lines represent the contraction of a primary overdensity with a secondary one. We define the ‘primary’ points with those that contain the second-order densities and depict them in the above diagram with two colored dots. We define the ‘secondary’ points with those that contain the first-order densities and depict them with one black dot. The above diagram does not show Wick contractions between the leftover-secondary densities (black dots), but what results from them has been explicitly included and calculated in each of the sections below. This figure visually illustrates that the partially-connected terms lack Wick contractions between all densities of the two tetrahedrons.}
\label{fig:2nd_pc_terms}
\end{figure}

As demonstrated in Appendix \ref{sec:Form_4PCF_Cov}, the 1-loop covariance introduces partially-connected (PC) terms. There are two possible partially-connected configurations as shown in Figure~\ref{fig:2nd_pc_terms}. These arise when one of the second-order (primary) densities is Wick-contracted with a linear (secondary) density inside the same tetrahedron, forming an internal loop. As a result, the connection between the two tetrahedra is incomplete, in contrast to the fully-connected case. The two possible structures are labeled PC–Primary–Secondary and PC–Half–Half, and we evaluate them in this order. In the diagrams, solid lines again denote contractions between densities of the same order, while dotted lines indicate contractions between primary and secondary fields. Together, these two cases exhaust all possible partially-connected structures at tenth order in the linear density field.

\subsubsection{PC-Primary-Secondary Configuration}
We begin the analysis of the PC-primary-secondary configuration as shown in Fig. \ref{fig:2nd_pc_terms}. Using Wick's theorem for the PC-primary-secondary configuration on Eq. (\ref{eq:sub_cov_before_Wicks}) results in:
\begin{align}
&{\rm Cov}_{4, \rm 1L}^{[2],{\rm PC-P-S}}(\mathbf{R},\mathbf{R}') = \prod_{v=0}^{3} \int_{\vecs}\int_{\veck_v}\int_{\veck'_v}\left[ \;e^{-i\veck_v\cdot(\vecx+\vecr_v)} e^{-i\veck'_v \cdot(\vecx+\vecr'_v+\vecs)} \right]\nonumber \\
& \qquad \qquad \qquad \times\int d^3\vecq_1 \int d^3\vecq_2 \int d^3\vecq'_1 \int d^3\vecq'_2 \;\W(\vecq_1,\vecq_2) \nonumber \\
& \qquad \qquad \qquad \times  \W(\vecq'_1,\vecq'_2)\; \dD(\veck_0-\vecq_1-\vecq_2)\; \dD(\veck_0'-\vecq'_1-\vecq'_2)\nonumber \\
& \qquad \qquad \qquad \times  \left<\widetilde{\delta}_{\rm lin.}(\vecq_1)\widetilde{\delta}_{\rm lin.}(\veck_1)\right>\left<\widetilde{\delta}_{\rm lin.}(\vecq_2)\widetilde{\delta}_{\rm lin.}(\veck_2)\right> \left<\widetilde{\delta}_{\rm lin.}(\vecq'_1)\widetilde{\delta}_{\rm lin.}(\veck'_1)\right>\nonumber \\
& \qquad \qquad \qquad \times\left<\widetilde{\delta}_{\rm lin.}(\vecq'_2)\widetilde{\delta}_{\rm lin.}(\veck'_2)\right> \left<\widetilde{\delta}_{\rm lin.}(\veck_3)\widetilde{\delta}_{\rm lin.}(\veck'_3)\right> + 35\; {\rm perm.},
\end{align}
where $35$ perm. indicates the other 35 combinations\footnote{Finding the number of permutations is done in four steps. i) Choose one of the second-order densities in either of the two tetrahedra, then pair this second-order density with any linear density from the same tetrahedron; one has 3 possibilities. ii) Pair the remaining second-order density with any of the linear densities in the same tetrahedron; one has 2 possibilities. iii) Repeat i) and ii) in the other tetrahedron; one has 6 possibilities. Total number of permutations = 3 $\times$ 2$\times$ 6 = 36 perms.} of Wick's theorem that we can obtain for this structure. For brevity, we leave this out, but it is implicitly included into our calculations. Evaluating the ensemble average as power spectra and performing the integrals $\vecq_s$, $\vecq'_s$, $\veck_0$, $\veck'_0$ and $\veck_3$ by taking advantage of the Dirac delta functions, we obtain:
\begin{align}
&{\rm Cov}_{4, \rm 1L}^{[2],{\rm PC-P-S}}(\mathbf{R},\mathbf{R}') = (2\pi)^{15} \int_{\vecs} I^{(3)}(\vecr_3,-\vecs,-\vecr'_3) \nonumber \\
& \qquad \qquad \qquad \times \int_{\veck_1}e^{-i\veck_1\cdot(\vecr_1 - \vecr_0)}P_{\rm lin}(k_1)\nonumber \\
& \qquad \qquad \qquad \times\int_{\veck'_1}e^{-i\veck'_1\cdot(\vecr'_1 -\vecr'_0)}P_{\rm lin}(k'_1)\nonumber \\
& \qquad \qquad \qquad \times \int_{\veck_2} \W(\veck_1,\veck_2)P_{\rm lin}(k_2)  e^{-i\veck_2\cdot(\vecr_2 -\vecr_0)}\nonumber \\
& \qquad \qquad \qquad \times \int_{\veck'_2}\W(\veck'_1,\veck'_2)P_{\rm lin}(k'_2)  e^{-i\veck'_2\cdot(\vecr'_2 -\vecr'_0)},
\end{align}
where we have used the same definition of Eq. (\ref{eq:I_PS_1}) on $I^{(3)}$; we changed the number in parentheses from $1\rightarrow3$ to represent the result from the integral of $\veck_3$.

We obtain the results for the remaining four integrals by expanding the exponential using the plane wave expansion, and the second-order kernel with Eq. (\ref{eq:W_2_def}):

\begin{align}
&I_{j,n}^{(2)}(\vecr_0,\vecr_2) = (4\pi) \sum_{L_{22},L_{20}=0}^{\infty}\sum_{M_{22}=-L_{22}}^{L_{22}} \sum_{M_{20}=-L_{20}}^{L_{20}} w_{L_{22},L_{20}} G^{L_{22},L_{20}, j}_{M_{22},M_{20},m_j} \nonumber \\
&  \qquad \qquad \qquad \times f^{[n]}_{L_{20},L_{22}}(r_0,r_2) Y_{L_{20},M_{20}}(\hatr_0)Y_{L_{22},M_{22}}^*(\hatr_2),
\end{align}
\begin{align}
&I_{j',n'}^{(2p)}(\vecr'_0,\vecr'_2) = (4\pi) \sum_{L'_{22},L'_{20}=0}^{\infty}\sum_{M'_{22}=-L'_{22}}^{L'_{22}} \sum_{M'_{20}=-L'_{20}}^{L'_{20}} w_{L'_{22},L'_{20}} G^{L'_{22},L'_{20},j'}_{M'_{22},M'_{20},m_{j'}}  \nonumber \\
&  \qquad \qquad \qquad \times f^{[n']}_{L'_{20},L'_{22}}(r_0,r_2)Y_{L'_{20},M'_{20}}(\hatr'_0)Y_{L'_{22},M'_{22}}^*(\hatr'_2),
\end{align}
\begin{align}
&I_{j,n}^{(1)}(\vecr_0,\vecr_1) = (4\pi) \sum_{L_{11},L_{10}=0}^{\infty}\sum_{M_{11}=-L_{11}}^{L_{11}} \sum_{M_{10}=-L_{10}}^{L_{10}} w_{L_{11},L_{10}} G^{L_{11},L_{10},j}_{M_{11},M_{10},m_j} \nonumber \\
&  \qquad \qquad \qquad \times f^{[-n]}_{L_{10},L_{11}}(r_0,r_1) Y_{L_{10},M_{10}}(\hatr_0)Y_{L_{11},M_{11}}^*(\hatr_1),
\end{align}
\begin{align}
&I_{j',n'}^{(1p)}(\vecr'_0,\vecr'_1) = (4\pi) \sum_{L'_{11},L'_{10}=0}^{\infty} \sum_{M'_{11}=-L'_{11}}^{L'_{11}} \sum_{M'_{10}=-L'_{10}}^{L'_{10}} w_{L'_{11},L'_{10}} G^{L'_{11},L'_{10},j'}_{M'_{11},M'_{10},m_{j'}} \nonumber \\
&  \qquad \qquad \qquad \times f^{[-n']}_{L'_{10},L'_{11}}(r'_0,r'_1) Y_{L'_{10},M'_{10}}(\hatr'_0)Y_{L'_{11},M'_{11}}^*(\hatr'_1),
\end{align}
where $G$ is the Gaunt coefficient, Eq. (\ref{eq:Gaunt_Coeff}), $w$ is:
\begin{align}\label{eq:w_constant_def}
    w_{L,L'}\equiv i^{L-L'},
\end{align}
and the radial integral is defined as:
\begin{align}\label{eq:f_radial_def}
f^{[n]}_{\ell,\ell'}(r,r')\equiv \int \frac{dk}{2\pi^2}\;k^{n+2}\;j_{\ell}(kr) j_{\ell'}(kr')\;P(k).
\end{align}

\begin{figure}[h!]
\centering
\includegraphics[scale=0.5]{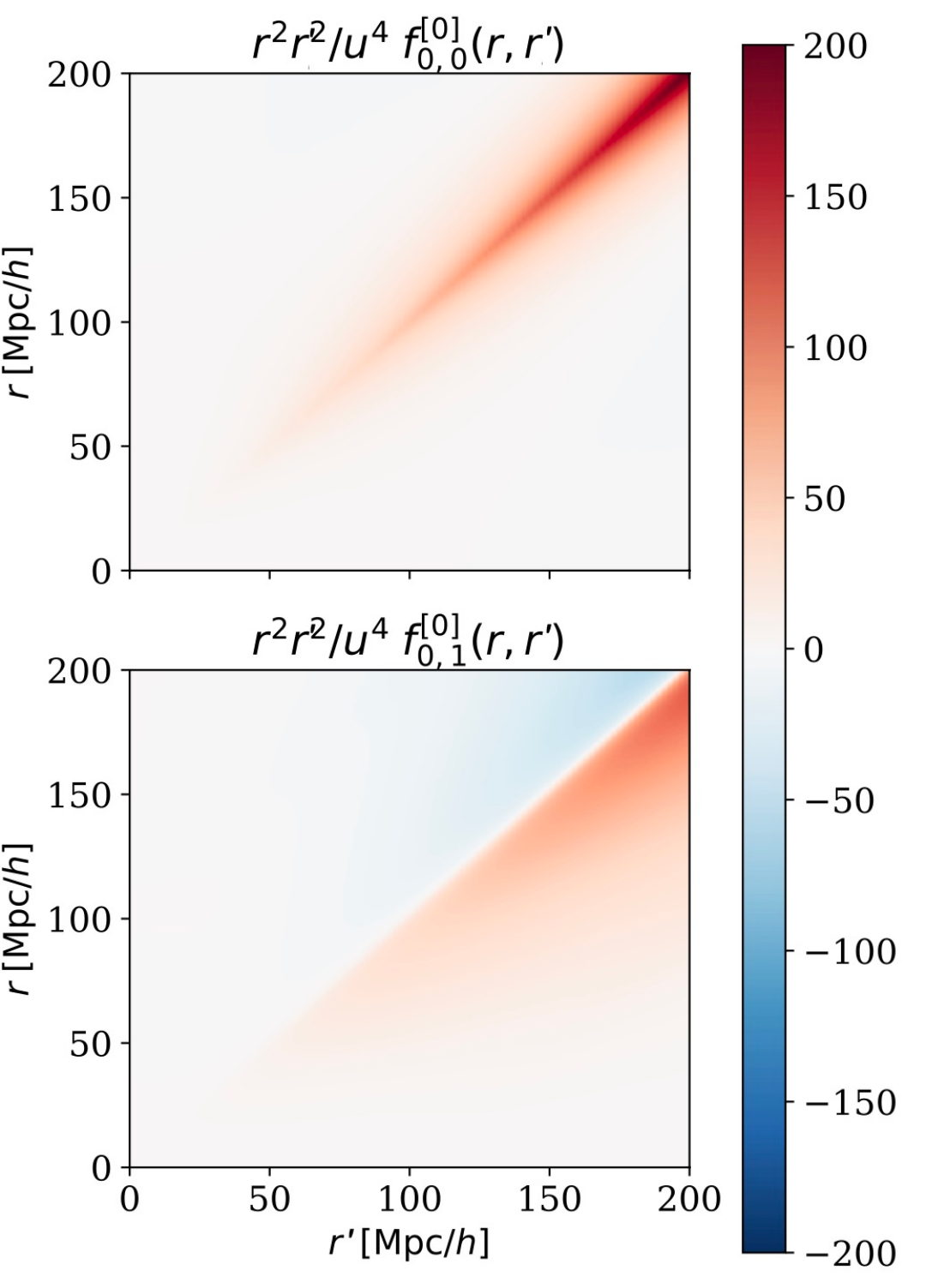}
\caption{Here, we show Eq. (\ref{eq:f_radial_def}) choosing $n = 0$, $\ell=0$, and $\ell' = \left\{0,1\right\}$. The plot has been taken from section 3 of \cite{Ortola_4PCF}. The \textit{upper panel} shows the integral for $\ell'=0$ and is largest along the diagonal. The \textit{lower panel} shows the integral for $\ell'=1$ and is largest in the off-diagonal elements. Since the 4PCF can be approximated as the square of the 2PCF on large scales, $(\xi_{0})^{2}(r) \sim (1/r^2)^2$, we have weighted the integral by $r^2r_i^2/u^{4}$, with $u \equiv 10 \;\left[{\rm Mpc}/h\right]$, to take out its fall-off.}
\label{fig:fint}
\end{figure}

After evaluation of the $\hats$ integral using Eq. (\ref{eq:s_hat_PC-PS_result}), we find: 
\begin{empheq}[box=\widefbox]{align}\label{eq:PC-PS_res}
&{\rm Cov}_{4, \rm 1L}^{[2],{\rm PC-P-S}}(\mathbf{R},\mathbf{R}') = (2\pi)^{15} (4\pi)^8 \sum_{\rm All}c_{j,n}^{(\mu)}c_{j',n'}^{(\beta)}\;\overline{\mathcal{Q}}_{(\rm PC-P-S)}^{\Lambda,\Lambda',M,M'}\nonumber \\
&\qquad \qquad \qquad \qquad \quad \times w_{L_{11},L_{10}} G^{L_{11},L_{10},j}_{M_{11},M_{10},m_j} \;w_{L_{22},L_{20}} G^{L_{22},L_{20}, j}_{M_{22},M_{20},m_j}\nonumber \\
&\qquad \qquad \qquad \qquad \quad \times w_{L'_{11},L'_{10}} G^{L'_{11},L'_{10},j'}_{M'_{11},M'_{10},m_{j'}}\; w_{L'_{22},L'_{20}} G^{L'_{22},L'_{20},j'}_{M'_{22},M'_{20},m_{j'}} \nonumber \\
&\qquad \qquad \qquad \qquad \quad \times C_{L_3,L_{3s},L'_{3}} \Upsilon_{L_3,L_{3s},L'_{3}} \mathcal{G}_{L_3,L_{3s},L'_{3}}\dK_{L_{3s},0}\dK_{L_{3},L'_3}\nonumber \\
&\qquad \qquad \qquad \qquad \quad \times f_{L_{10},L_{11}}^{[-n]} (r_0,r_1)f_{L'_{10},L'_{11}}^{[-n']} (r'_0,r'_1)\nonumber \\
&\qquad \qquad \qquad \qquad \quad \times f_{L_{20},L_{22}}^{[n]} (r_0,r_2)f_{L'_{20},L'_{22}}^{[n']} (r'_0,r'_2)\nonumber \\
&\qquad \qquad \qquad \qquad \quad \times S_{L_3,L'_3, \;{\rm (PC-P-S)}}^{(L_{3s})}(r_3,r'_3) \PP_{\Lambda}(\hatR)\PP_{\Lambda'}(\hatR'),
\end{empheq}
where the $\Lambda$ and $\Lambda'$ momenta have been defined in $\S$\ref{sec:s_hat_PC-PS_Evaluation}. Table \ref{table:1} provides the equation where all the constants have been defined. We have defined the radial integral $S$ as:
\begin{align}\label{eq:Sint_PC-PS}
&S_{L_3,L'_3, \;{\rm (PC-P-S)}}^{(L_{3s})}(r_3,r'_3) \equiv \int ds\;s^2\; g_{L_3,L_{3s},L'_{3}}^{[0]} (r_3,s,r'_3).
\end{align} 
The subscripts $L_3,L'_3$ denote the set of all angular momentum indices that affect the resulting variables (\textit{i.e.}, the $r_3$, $r'_3$ variables). The superscript $L_{3s}$ denotes the angular momentum index that couples to $s$ in the spherical Bessel function within the $g$ radial integral. 

\begin{figure}[h!]
\centering
\includegraphics[scale=0.6]{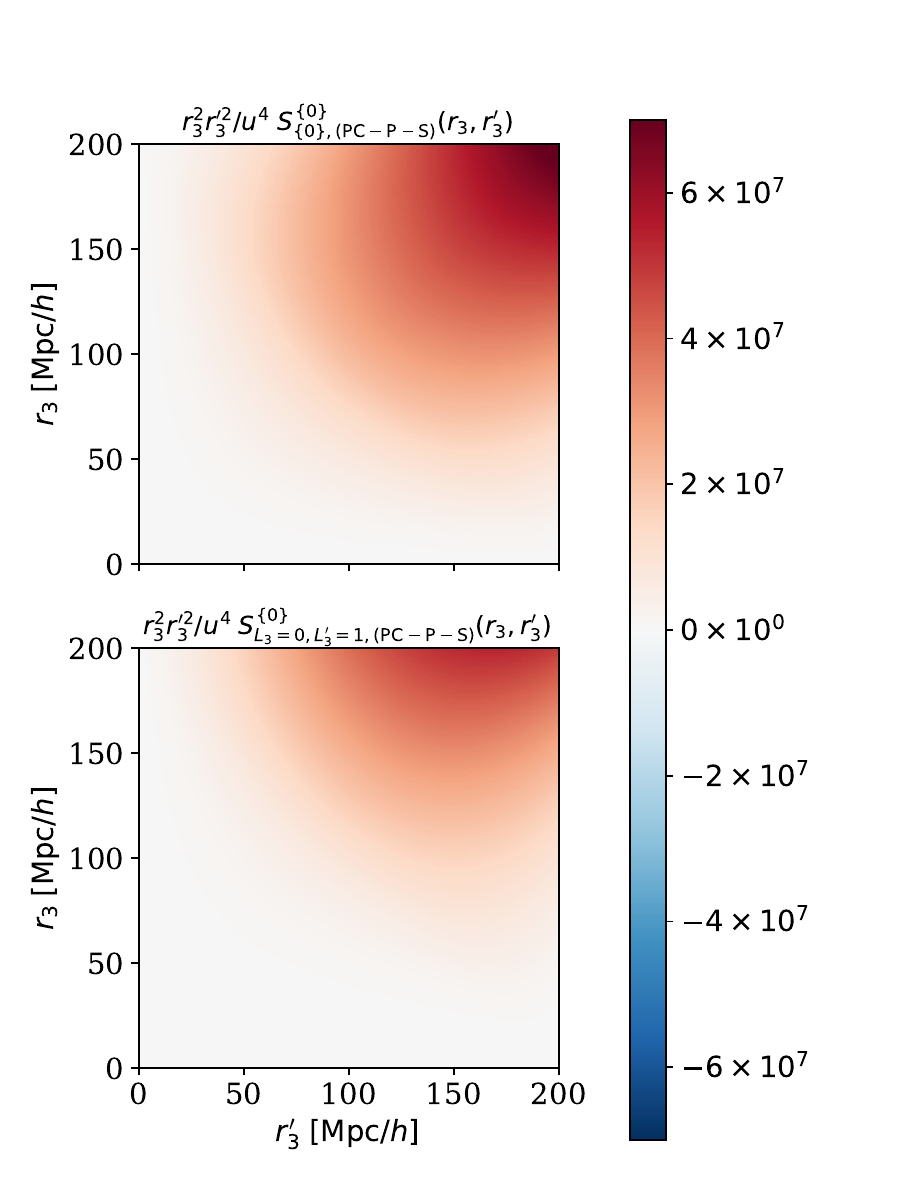}
\caption{Here, we show Eq. (\ref{eq:Sint_PC-PS}) choosing $n=0$, $L_3 = 0$, and $L_3' = \left\{0,1\right\}$. The \textit{upper panel} shows the integral for $L'_3=0$, while the \textit{lower panel} shows the integral for $L_3'=1$. Since the 4PCF can be approximated as the square of the 2PCF on large scales, $(\xi_{0})^{2}(r) \sim (1/r^2)^2$, we have weighted the integral by $r^2r'^2/u^{4}$, with $u \equiv 10 \;\left[{\rm Mpc}/h\right]$, to take out its fall-off. The $S$ integral is composed of the intermediate radial integrals, $g$, which reveals a rectangular boundary as shown in Fig. \ref{fig:gint2}. The integration of the different rectangular boundaries creates the oval-shaped regions shown in both panels. Analytical results for similar radial integrals with the power spectrum being a power law have been obtained in \cite{Ortola_4PCF}.}
\label{fig:PC_PS_int}
\end{figure}

\subsubsection{PC-Half-Half Configuration}
We finish with the analysis of the PC-half-half configuration as shown in Fig. \ref{fig:2nd_pc_terms}. Using Wick's theorem for the PC-half-half configuration on Eq. (\ref{eq:sub_cov_before_Wicks}) results in:
\begin{align}
&{\rm Cov}_{4, \rm 1L}^{[2],{\rm PC-H-H}}(\mathbf{R},\mathbf{R}') = \prod_{v=0}^{3} \int_{\vecs}\int_{\veck_v}\int_{\veck'_v}\left[ \;e^{-i\veck_v\cdot(\vecx+\vecr_v)} e^{-i\veck'_v \cdot(\vecx+\vecr'_v+\vecs)} \right]\nonumber \\
& \qquad \qquad \qquad \times\int d^3\vecq_1 \int d^3\vecq_2 \int d^3\vecq'_1 \int d^3\vecq'_2 \;\W(\vecq_1,\vecq_2) \nonumber \\
& \qquad \qquad \qquad \times  \W(\vecq'_1,\vecq'_2)\; \dD(\veck_0-\vecq_1-\vecq_2)\; \dD(\veck_0'-\vecq'_1-\vecq'_2)\nonumber \\
& \qquad \qquad \qquad \times  \left<\widetilde{\delta}_{\rm lin.}(\vecq_1)\widetilde{\delta}_{\rm lin.}(\veck_1)\right>\left<\widetilde{\delta}_{\rm lin.}(\vecq_2)\widetilde{\delta}_{\rm lin.}(\vecq'_2)\right> \left<\widetilde{\delta}_{\rm lin.}(\vecq'_1)\widetilde{\delta}_{\rm lin.}(\veck'_1)\right>\nonumber \\
& \qquad \qquad \qquad \times\left<\widetilde{\delta}_{\rm lin.}(\veck_2)\widetilde{\delta}_{\rm lin.}(\veck'_2)\right> \left<\widetilde{\delta}_{\rm lin.}(\veck_3)\widetilde{\delta}_{\rm lin.}(\veck'_3)\right> + 35\; {\rm perm.},
\end{align}
where $35$ perm. indicates the other 35 combinations\footnote{Finding the number of permutations is done in four steps: i) choose one of the second-order densities in any of the two tetrahedra, then pair this second-order density with any linear densities from the same tetrahedron; one has 3 possibilities. ii) Pair the remaining second-order density with any of the second-order densities in the other tetrahedron; one has 2 possibilities. iii) with the remaining second-order density in the other tetrahedron Repeat i); one has 3 possibilities. iv) choose one of the linear densities in any of the two tetrahedra, then pair this density with a linear density in the other tetrahedron; one has 2 possibilities. Total number of permutations = 3 $\times$ 2$\times$ 3 $\times$ 2  = 36 perms.} of Wick's theorem that we can obtain for this structure. For brevity, we leave this out, but it is implicitly included in our calculations. 

Evaluating the ensemble averages as power spectra and performing the integrals the $\vecq_s$ (except for $\vecq_2$), $\vecq'_s$, $\veck_0$, $\veck'_0$, $\veck_2$ and $\veck_3$ by taking advantage of the Dirac delta functions, we obtain:
\begin{align}
&{\rm Cov}_{4, \rm 1L}^{[2],{\rm PC-H-H}}(\mathbf{R},\mathbf{R}') = (2\pi)^{15} \int_{\vecs} I^{(2)}(\vecr_2,-\vecs,-\vecr'_2) I^{(3)}(\vecr_3,-\vecs,-\vecr'_3) \nonumber \\
& \qquad \qquad \qquad \times \int_{\vecq_2}e^{-i\vecq_2\cdot(\vecr_0 -\vecs- \vecr'_0)}P_{\rm lin}(q_2)\nonumber \\
& \qquad \qquad \qquad \times \int_{\veck_1} \W(-\veck_1,\vecq_2)P_{\rm lin}(k_1)  e^{-i\veck_1\cdot(\vecr_1 -\vecr_0)}\nonumber \\
& \qquad \qquad \qquad \times \int_{\veck'_1}\W(\veck'_1,\vecq_2)P_{\rm lin}(k'_1)  e^{-i\veck'_1\cdot(\vecr'_1 -\vecr'_0)},
\end{align}
where we have used the same definition of Eq. (\ref{eq:I_PS_1}) on $I^{(2)}$ and $I^{(3)}$; we changed the number in parentheses from $1\rightarrow2$ and $1\rightarrow3$ to represent the result from the integrals of $\veck_2$ and $\veck_3$. 

We obtain the result of the remaining three integrals by expanding the exponential using the plane wave expansion, and the second-order kernel with Eq. (\ref{eq:W_2_def}):
\begin{align}
&I_{j,n}^{(1)}(\vecr_0,\vecr_2) = (4\pi) \sum_{L_{11},L_{10}=0}^{\infty}\sum_{M_{11}=-L_{11}}^{L_{11}}\sum_{M_{10}=-L_{10}}^{L_{10}} w_{L_{11},L_{10}} G^{L_{11},L_{10},j}_{M_{11},M_{10},m_j} \nonumber\\
&\qquad \qquad \qquad \times f^{[n]}_{L_{10},L_{11}}(r_0,r_1) Y_{L_{10},M_{10}}(\hatr_0)Y_{L_{11},M_{11}}^*(\hatr_1),
\end{align}
\begin{align}
&I_{j',n'}^{(1p)}(\vecr'_0,\vecr'_1) = (4\pi)  \sum_{L'_{11},L'_{10}=0}^{\infty}\sum_{M'_{11}=-L'_{11}}^{L'_{11}}\sum_{M'_{10}=-L'_{10}}^{L'_{10}}  w_{L'_{11},L'_{10}} G^{L'_{11},L'_{10},j'}_{M'_{11},M'_{10},m_{j'}} \nonumber\\
&\qquad \qquad \qquad \times  f^{[n']}_{L'_{10},L'_{11}}(r'_0,r'_1) Y_{L'_{10},M'_{10}}(\hatr'_0)Y_{L'_{11},M'_{11}}(\hatr'_1),
\end{align}
\begin{align}
&I_{j,j',n,n'}^{(q_2)}(\vecr_0,\vecr_1) = (4\pi)^2 \sum_{L_{q20},L_{q2s},L'_{q20}} C_{L_{q20},L_{q2s},L'_{q20}} \Upsilon_{L_{q20},L_{q2s},L'_{q20}} \mathcal{J}_{L_{q20},L_{q2s},L'_{q20}}^{j,j',m_j,m_{j'}}\nonumber\\
& \qquad \qquad \qquad\qquad \;\;\;\;\times g^{[-n-n']}_{L_{q20},L_{q2s},L'_{q20}}(r_0,s,r'_0) \PP_{L_{q20},L_{q2s},L'_{q20}}(\hatr_0,-\hats,-\hatr'_0).
\end{align}
After evaluation of the $\hats$ integral using Eq. (\ref{eq:Q_bar_def_PC-HH}), we find: 
\begin{empheq}[box=\widefbox]{align}\label{eq:PC-HH_res}
&{\rm Cov}_{4, \rm 1L}^{[2],{\rm PC-H-H}}(\mathbf{R},\mathbf{R}') = (2\pi)^{15} (4\pi)^{10}\sum_{\rm All}c_{j,n}^{(\mu)}c_{j',n'}^{(\beta)}\;\overline{\mathcal{Q}}_{(\rm PC-H-H)}^{\Lambda,\Lambda_s,\Lambda',\bar{M}}\nonumber \\
&\qquad \qquad \qquad \qquad \quad \times w_{L_{11},L_{10}} G^{L_{11},L_{10},j}_{M_{11},M_{10},m_j} \;w_{L'_{11},L'_{10}} G^{L'_{11},L'_{10},j'}_{M'_{11},M'_{10},m_{j'}}\nonumber \\
&\qquad \qquad \qquad \qquad \quad \times C_{L_{q20},L_{q2s},L'_{q20}} \Upsilon_{L_{q20},L_{q2s},L'_{q20}} \mathcal{J}_{L_{q20},L_{q2s},L'_{q20}}^{j,j',m_j,m_{j'}}\nonumber \\
&\qquad \qquad \qquad \qquad \quad \times C_{L_2,L_{2s},L'_{2}} \Upsilon_{L_2,L_{2s},L'_{2}} \mathcal{G}_{L_2,L_{2s},L'_{2}}\nonumber \\
&\qquad \qquad \qquad \qquad \quad \times C_{L_3,L_{3s},L'_{3}} \Upsilon_{L_3,L_{3s},L'_{3}} \mathcal{G}_{L_3,L_{3s},L'_{3}}\nonumber \\
&\qquad \qquad \qquad \qquad \quad \times f^{[n]}_{L_{10},L_{10}}(r_0,r_1) f^{[n']}_{L'_{10},L'_{10}}(r'_0,r'_1) \nonumber \\
&\qquad \qquad \qquad \qquad \quad \times S_{\{L\}, \;{\rm (PC-H-H)}}^{(\{L_{is}\})}(r_0,r'_0,r_2,r'_2,r_3,r'_3) \PP_{\Lambda}(\hatR)\PP_{\Lambda'}(\hatR'),
\end{empheq}
where the $\Lambda,\; \Lambda_s$, $\Lambda'$, and $\bar{M}$ momenta have been defined in $\S$\ref{sec:s_hat_PC-HH_Evaluation}. Table \ref{table:1} provides the equation where all the constants have been defined. We have defined the radial integral $S$ as:
\begin{align}\label{eq:Sint_PC_HH}
&S_{\{L\}, \;{\rm (PC-H-H)}}^{(\{L_{is}\})}(r_0,r'_0,r_2,r'_2,r_3,r'_3) \equiv \int ds\;s^2\; g^{[-n-n']}_{L_{q20},L_{q2s},L'_{q20}}(r_0,s,r'_0)\nonumber \\ & \qquad \qquad \qquad \qquad \qquad\qquad \qquad \qquad \times g^{[0]}_{L_{2},L_{2s},L'_{2}}(r_2,s,r'_2)g^{[0]}_{L_{3},L_{3s},L'_{3}}(r_3,s,r'_3).
\end{align} 
The subscript $\{L\}$ denotes the set of all angular momentum indices that affect one of the resulting variables (\textit{i.e.}, the $r_0$, $r'_0$,$\cdots$ variables). The superscript $(\{L_{is}\})$ denotes the set of all angular momentum indices that are coupled to $s$ in the spherical Bessel functions.  

\begin{figure}[h!]
\centering
\includegraphics[scale=0.6]{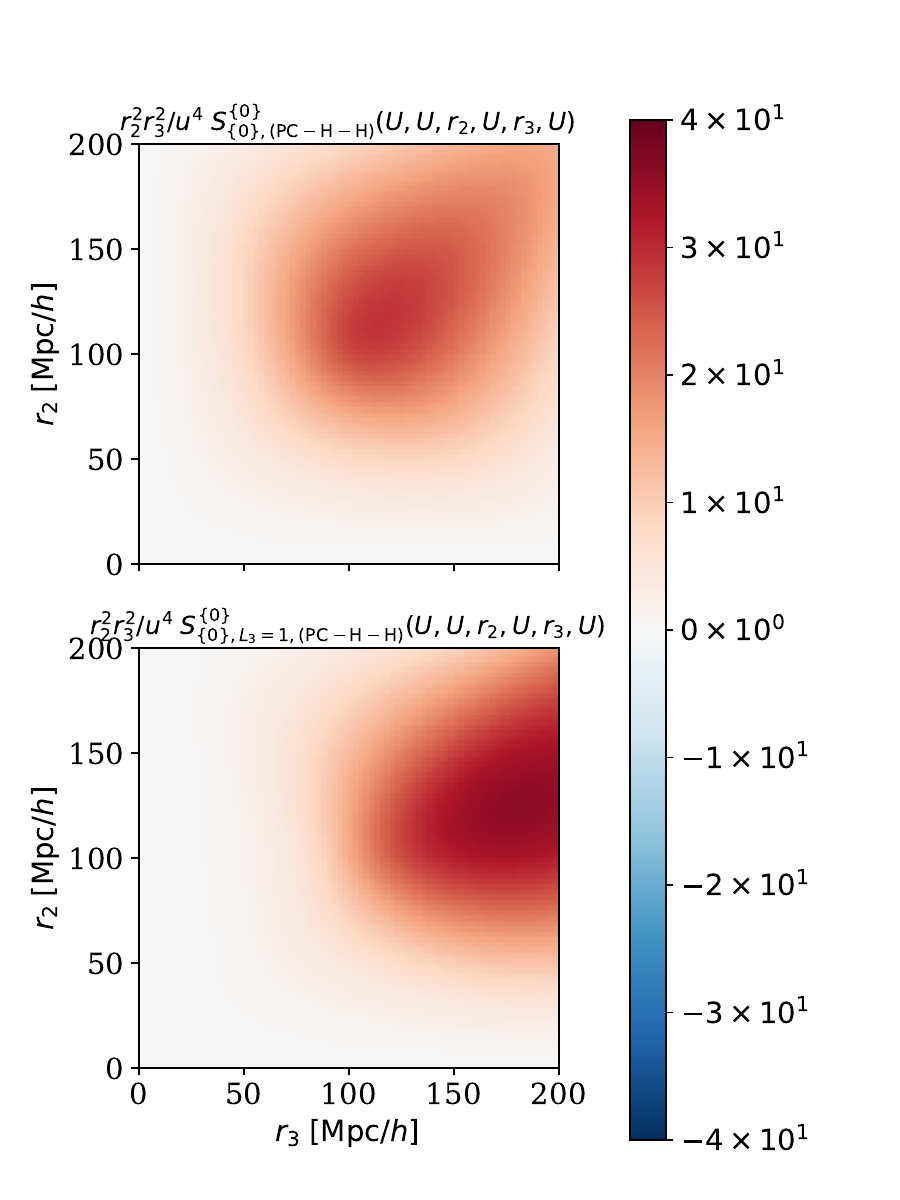}
\caption{Here, we show Eq. (\ref{eq:Sint_PC_HH}) choosing $n=n'=0$ and all angular momenta $\{L\} = 0$ except $L_3 = \left\{0,1\right\}$. The \textit{upper panel} shows the integral for $L_3=0$, while the \textit{lower panel} shows the integral for $L_3=1$. Since the 4PCF can be approximated as the square of the 2PCF on large scales, $(\xi_{0})^{2}(r) \sim (1/r^2)^2$, we have weighted the integral by $r^2r'^2/u^{4}$, with $u \equiv 10 \;\left[{\rm Mpc}/h\right]$, to take out its fall-off. The $S$ integral is composed of the intermediate radial integrals, $g$, which reveals a rectangular boundary as shown in Fig. \ref{fig:gint2}. The integration of the different rectangular boundaries creates the oval-shaped regions shown in both panels. Analytical results for similar radial integrals with the power spectrum being a power law have been obtained in \cite{Ortola_4PCF}.}
\label{fig:PC_PS_int}
\end{figure}

\section{Discussion and Conclusions}
In this work, we have developed a systematic framework to compute the covariance of the 4PCF beyond the Gaussian approximation using the second-order density contrast and galaxy biasing. The key step is to generalize all second-order kernels into a single effective kernel, $W^{(2)}$, defined in Eq. (\ref{eq:W_2_def}), which allows us to treat the second-order contributions, both from the density and from galaxy biasing, in a unified way. Applying Wick’s theorem to the density contrasts, we find that all contractions can be organized into five configurations: three fully-connected cases, shown in Eqs. (\ref{eq:PS_res}), (\ref{eq:HH_res}), and (\ref{eq:PP_result}), and two partially-connected cases, given in Eqs. (\ref{eq:PC-PS_res}) and (\ref{eq:PC-HH_res}). These are illustrated in Figs. \ref{fig:dc_configurations} and \ref{fig:2nd_pc_terms}.

The practical implication is that the covariance due to second-order densities can be built from these five structures alone, with the specific perturbative kernel entering only through the generalized form of $W^{(2)}$. This provides a direct procedure for constructing the full second-order contribution to the 4PCF covariance, while the third-order density contribution is computed in our companion paper \cite{Ortola_4PCF_Cov_II}. 

In the Gaussian treatment, the covariance is reduced to products of two-point functions through Wick's theorem, whereas here we explicitly compute the NLO terms sourced by second-order densities. These provide corrections to the Gaussian covariance that are expected to become increasingly relevant on mildly nonlinear scales, as shown in Appendix \ref{sec:Correction_Scale}, where the Gaussian approximation becomes incomplete.

At the same time, the present calculation has a number of limitations. Due to the increased complexity of the calculation, we have not included redshift-space distortions, survey window effects, or shot noise. In addition, our treatment is based on standard perturbation theory, and we do not include ultraviolet counterterms or infrared resummation. The results presented here should therefore be understood as a first step toward a more complete 4PCF covariance model.

Since this framework must ultimately be useful in practice, it is also important to assess both its numerical cost and the stability of the resulting sums. For this reason, we evaluate the computational cost of our approach in Appendix \ref{sec:Computational_Cost} and study the convergence of the structures in Appendix \ref{sec:Convergence_Test}.

Looking ahead, important next steps include extending this framework to incorporate redshift-space distortions and survey effects. We will also perform numerical validations against $N$-body simulations and mock catalogs to assess the accuracy and domain of validity of our perturbative treatment. These developments will allow us to provide survey-ready analytical covariance matrices for upcoming data releases from DESI and beyond. 

We have also made available a public \href{https://doi.org/10.5281/zenodo.20528953}{Zenodo repository} containing code to reproduce selected figures from this paper.

\acknowledgments
We thank the Slepian research group for useful discussions and comments on this work. ZS especially thanks Bob Cahn for useful discussions. This publication was made possible through the support of Grant 63041 from the John Templeton Foundation. The opinions expressed in this publication are those of the author(s) and do not necessarily reflect the views of the John Templeton Foundation. ZS acknowledges funding from NASA grant number 80NSSC24M0021 and funding from UF Research AI award \#00133699. The authors acknowledge UFIT Research Computing for providing computational resources on HiPerGator and support that have contributed to the research results reported in this publication.

\appendix

\section{Generalizing the Second-Order Density Contrasts for Use in the Covariance Matrix}\label{sec:Generalization_delta_2}
In this section we generalize $\delta^{(2)},\; \delta_{\rm lin.}^2$ and $S^{(2)}$ to a single second-order density contrast in Fourier space. We do so using a generalized second-order kernel we term $W^{(2)}$, defined in Eq. (\ref{eq:W_2_def}). 

\subsection{Second-Order Density Kernel}
We begin by expressing the second-order density contrast using its inverse Fourier transform:
\begin{align}\label{eq:delta_2_def_FS}
&\delta^{(2)}(\vecx + \vecr_i) = \int \frac{d^{3}\veck_i}{(2\pi)^3} \;e^{-i \veck_i\cdot (\vecx+\vecr_i)} \widetilde{\delta}^{(2)}(\veck_i) \nonumber \\ 
& = \int \frac{d^{3}\veck_i}{(2\pi)^3} \;e^{-i \veck_i\cdot (\vecx+\vecr_i)} \int d^3\vecq_1 \;d^3\vecq_2\; \dD(\veck_i - \vecq_1 - \vecq_2)\nonumber \\ 
& \qquad\times F^{(2)}(\vecq_1,\vecq_2) \;\widetilde{\delta}_{\rm lin.}(\vecq_1) \widetilde{\delta}_{\rm lin.}(\vecq_2) \nonumber\\
& = \int d^3\vecq_1 \;d^3\vecq_2 \;e^{-i (\vecq_1+\vecq_2)\cdot (\vecx+\vecr_i)} F^{(2)}(\vecq_1,\vecq_2) \;\widetilde{\delta}_{\rm lin.}(\vecq_1) \widetilde{\delta}_{\rm lin.}(\vecq_2).
\end{align}
To obtain the second equality, we used the definition of the second-order density contrast in terms of the symmetrized second-order kernel $F^{(2)}$ \cite{Bernardeau}. To obtain the third equality, we evaluated the Dirac delta function to perform the $\veck_i$ integral. $F^{(2)}$ is defined as \cite{Farshad, Ortola_4PCF}:
\begin{align}\label{eq:F2_def}
&F^{(2)}\left(\mathbf{k}_2,\mathbf{k}_3\right) = \frac{17}{21} \mathcal{L}_{0}(\mathbf{\widehat{k}}_2 \cdot \mathbf{\widehat{k}}_3) + \frac{1}{2} \left(\frac{k_2}{k_3}+\frac{k_3}{k_2}\right)\mathcal{L}_{1}(\mathbf{\widehat{k}}_2 \cdot \mathbf{\widehat{k}}_3) + \frac{4}{21}\mathcal{L}_{2}(\mathbf{\widehat{k}}_2 \cdot \mathbf{\widehat{k}}_3)\nonumber \\
&\qquad \qquad \quad\;\;=4\pi\sum_{j=0}^2\sum_{n=-1}^{1} c_{j,n}^{(F)} \;\mathcal{P}_{j}(\hatk_2,\hatk_3)\; k_2^n k_3^{-n}.
\end{align}
In the first line, $\mathcal{L}_{\ell}$ refers to the Legendre polynomial of order $\ell$. In the last equality we have expressed the kernel in terms of the isotropic basis functions as already derived in \cite{Ortola_4PCF}. The values of $c^{(F)}_{j,n}$ are $c^{(F)}_{0,0}=17/21$, $c^{(F)}_{1,1}=c^{(F)}_{1,-1}=-\sqrt{3}/2$, and $c^{(F)}_{2,0} = 4\sqrt{5}/21$; the rest of constants are zero. 

\subsection{Squared Linear Density Contrast}
We will now show that $\widetilde{\delta}_{\rm lin.}^2$ and $S^{(2)}$ can be expressed in the same form but with different coefficients. We first proceed with $\widetilde{\delta}_{\rm lin.}^2$:
\begin{align}\label{eq:delta_1_sq_def}
&\delta_{\rm lin.}^2(\vecx + \vecr_i) = \delta_{\rm lin.}(\vecx + \vecr_i) \delta_{\rm lin.}(\vecx + \vecr_i) \nonumber \\
& = \int \frac{d^{3}\veck_i}{(2\pi)^3}\; e^{-i \veck_i\cdot (\vecx+\vecr_i)} \widetilde{\delta}_{\rm lin.}(\veck_i)  \int \frac{d^{3}\veck''_i}{(2\pi)^3} \;e^{-i \veck''_i\cdot (\vecx+\vecr_i)}\widetilde{\delta}_{\rm lin.}(\veck''_i) \nonumber \\
& = \int \frac{d^{3}\veck_i\;d^{3}\veck''_i}{(2\pi)^6}\; e^{-i (\veck_i+\veck''_i)\cdot (\vecx+\vecr_i)} \widetilde{\delta}_{\rm lin.}(\veck_i) \widetilde{\delta}_{\rm lin.}(\veck''_i) \nonumber\\
& = (2\pi)^{6} \lim_{F^{(2)}\to 1} \delta^{(2)}(\vecx + \vecr_0).
\end{align}
In the third equality, we observed that the linear density squared is simply what would result if we took the limit that $F^{(2)}\to 1$ in our usual computation of the second-order density by comparing with Eq. (\ref{eq:delta_2_def_FS}).

\subsection{Tidal Tensor Kernel}
Continuing with the Tidal tensor kernel, we find in Fourier space:
\begin{align}\label{eq:S2_def}
&S^{(2)}(\vecx + \vecr_i) = \int \frac{d^{3}\veck_i}{(2\pi)^3} \;e^{-i \veck_i\cdot (\vecx+\vecr_i)}\int d^3\vecq_1 \;d^3\vecq_2\; \dD(\veck_i - \vecq_1 - \vecq_2)\nonumber \\ 
& \qquad \qquad\qquad\qquad\times S^{(2)}(\vecq_1,\vecq_2) \;\widetilde{\delta}_{\rm lin.}(\vecq_1) \widetilde{\delta}_{\rm lin.}(\vecq_2) \nonumber \\
& \qquad \qquad \quad\;\; =  \lim_{F^{(2)}\to 2/3\mathcal{L}_{2}(\vecq_1,\vecq_2)} \delta^{(2)}(\vecx+\vecr_i), 
\end{align}
with:
\begin{align}
S^{(2)}(\vecq_1,\vecq_2) = \frac{(\vecq_1\cdot\vecq_2)^2}{q_1^2q_2^2} - \frac{1}{3} = \frac{2}{3}\mathcal{L}_2(\hatq_1\cdot\hatq_2).
\end{align}
In Eq. \ref{eq:S2_def}, we observed that the Tidal tensor kernel is simply what would result if we took the limit that $F^{(2)}\to (2/3) \mathcal{L}_2$ in our usual computation of the second-order density by comparing with Eq. (\ref{eq:delta_2_def_FS}).

Eqs. (\ref{eq:delta_2_def_FS}), (\ref{eq:delta_1_sq_def}), and (\ref{eq:S2_def}) suggest that we can encapsulate their information in Fourier space in a single form for the second-order density contrast. We only need a new general second-order kernel that can be expressed in terms of any of these three second-order density contrasts. 
\subsection{Generalized Second-Order Density Contrast}
Below, we present the generalized second-order density contrast and kernel:
\begin{align}\label{eq:delta_i_def}
&\delta_{\mu}^{(2)}(\vecx+\vecr_0) = \int \frac{d^{3}\veck_0}{(2\pi)^3}\; e^{-i \veck_0\cdot (\vecx+\vecr_0)}\int d^3\vecq_1 \;d^3\vecq_2\; \dD(\veck_0 - \vecq_1 - \vecq_2)\nonumber \\ 
& \qquad \qquad\qquad\qquad\times W_{\mu}^{(2)}(\vecq_1,\vecq_2) \;\widetilde{\delta}_{\rm lin.}(\vecq_1) \widetilde{\delta}_{\rm lin.}(\vecq_2), 
\end{align}
with
\begin{align}\label{eq:W_2_def}
&W_{\mu}^{(2)}(\vecq_1,\vecq_2) = 4\pi\sum_{j=0}^2 \sum_{n=-1}^{1} c_{j,n}^{(\mu)} \;\mathcal{P}_{j}(\hatq_1,\hatq_2) q_1^n q_2^{-n} \nonumber \\
& \qquad \qquad \quad\;= 4\pi\sum_{j, m_j,n} c_{j,n}^{(\mu)} \;D_j \; Y_{j,m_j}(\hatq_1) Y_{j,m_j}^*(\hatq_2)\; q_1^n q_2^{-n}, 
\end{align}
where $\mu$ runs over $\{0,1,2\}$ such that $\left\{\widetilde{\delta}_0^{(2)},\widetilde{\delta}_1^{(2)},\widetilde{\delta}_2^{(2)}\right\} = \left\{S^{(2)}, \widetilde{\delta}^{(2)}, \widetilde{\delta}_{\rm lin.}^2\right\}$. If we evaluate $W^{(2)}  = S^{(2)}$, the only non-zero coefficient will be $c_{j,n}^{(\mu)} = c_{2,0}^{(0)} = 2(\sqrt{5}/3)$. If $W^{(2)}  = \widetilde{\delta}_{\rm lin.}^2$, the only non-zero coefficient will be $c_{j,n}^{(\mu)} = c_{0,0}^{(2)} = 1$. When $W^{(2)} = \delta^{(2)}$ the non-zero coefficients are $c^{(F)}_{0,0}=17/21$, $c^{(F)}_{1,1}=c^{(F)}_{1,-1}=-\sqrt{3}/2$, and $c^{(F)}_{2,0} = 4\sqrt{5}/21$ \cite{Ortola_4PCF}. In the second equality above we have defined $D_j\equiv(-1)^j/\sqrt{2j+1}$ \cite{Cahn_Iso}.

\section{Evaluation of the Angular Integrals for the Covariance with Second-Order Densities}\label{Sec:Const_Calculations}
 In this section we show the evaluation of the main angular integrals in this work. We start with the evaluation of one 3-argument isotropic basis function when all its arguments are the same: 
\begin{align}\label{eq:CalG}
&\mathcal{G}_{L_1,L_2,L_3} \equiv \int_{\hatk} \; \PP_{L_1,L_2,L_3}(\hatk,\hatk,\hatk)  \nonumber \\
&= (-1)^{L_1+L_2+L_3}\sum_{M_1,M_2,M_3}\tj{L_1}{L_2}{L_3}{M_1}{M_2}{M_3}\int_{\hatk} \;Y_{L_1,M_1}(\hatk)Y_{L_2,M_2}(\hatk)Y_{L_3,M_3}(\hatk) \nonumber \\
& = (-1)^{L_1+L_2+L_3}\left(\frac{(2L_1+1)(2L_2+1)(2L_3+1)}{4\pi}\right)^{1/2}\tj{L_1}{L_2}{L_3}{0}{0}{0}\sum_{M_1,M_2,M_3}\tj{L_1}{L_2}{L_3}{M_1}{M_2}{M_3}^2\nonumber \\
& = \left(\frac{(2L_1+1)(2L_2+1)(2L_3+1)}{4\pi}\right)^{1/2}\tj{L_1}{L_2}{L_3}{0}{0}{0}.
\end{align}
In the second equality, we have expanded the 3-argument isotropic basis function using its definition:
\begin{align}\label{eq:def_Iso_Basis}
&\mathcal{P}_{L,L',L^{''}}(\hatk,\hatk,\hatk) = (-1)^{L+L'+L^{''}}\sum_{M,M',M^{''}} \begin{pmatrix}
L & L' & L^{''}\\
M & M' & M^{''}
\end{pmatrix}\nonumber \\
& \qquad \qquad \qquad \qquad\times\int_{\hatk} Y_{L,M}(\hatk)  Y_{L',M'}(\hatk)  Y_{L^{''},M^{''}}(\hatk). 
\end{align}
In the third equality, we used the Gaunt integral of three spherical harmonics, given by \cite{DLMF} in their Eq. (34.3.22):
\begin{align}\label{eq:Gaunt_Coeff}
   G_{M_1,M_2,M_3}^{L_1,L_2,L_3}\equiv \left(\frac{(2L_1+1)(2L_2+1)(2L_3+1)}{4\pi}\right)^{1/2}\tj{L_1}{L_2}{L_3}{0}{0}{0}\tj{L_1}{L_2}{L_3}{M_1}{M_2}{M_3},
\end{align}
and in the fourth equality we used Eq. (34.3.18) in \cite{DLMF}, and dropped the negative sign since the $3j$ symbol imposes an even sum of $L_1,\;L_2$ and $L_3$. 

Next, we evaluate the integral of one 3-argument isotropic basis function, with all its arguments being the same, against one spherical harmonic of the same argument: 
\begin{align}\label{eq:Fancy_H_def}
&\mathcal{H}_{L,L',L^{''}}^{\ell,m} \equiv \int_{\hatk} \mathcal{P}_{L,L',L^{''}}(\hatk,\hatk,\hatk) Y_{\ell,m}(\hatk)  \nonumber \\
& \qquad \qquad  = (-1)^{L+L'+L^{''}}\sum_{M,M',M^{''}} \begin{pmatrix}
L & L' & L^{''}\\
M & M' & M^{''}
\end{pmatrix} \nonumber \\
& \qquad \qquad \qquad \times \int_{\hatk} Y_{L,M}(\hatk)  Y_{L',M'}(\hatk)  Y_{L^{''},M^{''}}(\hatk)  Y_{\ell,m}(\hatk) \nonumber \\
& \qquad \qquad   = (-1)^{L+L'+L^{''}} \sum_{M,M',M^{''}}\;\sum_{\ell',m'} \begin{pmatrix}
L & L' & L^{''}\\
M & M' & M^{''}
\end{pmatrix}\nonumber \\
& \qquad \qquad \qquad \times G_{L^{''}, \ell, \ell'}^{M^{''}m,-m'} G_{L, L', \ell'}^{M,M',m'}.
\end{align}
To obtain the result in the third equality, we first combined a product of two spherical harmonics into a sum over single ones, and then evaluated the integral of a product of three spherical harmonics as defined in Eq. (\ref{eq:Gaunt_Coeff}). 

Last, we evaluate the integral of a product of one 3-argument isotropic basis function whose arguments are all the same against two spherical harmonics:
\begin{align}\label{eq:Fancy_J_def}
&\mathcal{J}_{L,L',L^{''}}^{\ell,\ell^{'},m,m'} \equiv \int_{\hatk} \mathcal{P}_{L,L',L^{''}}(\hatk,\hatk,\hatk) Y_{\ell,m}(\hatk) Y_{\ell',m'}(\hatk)  \nonumber \\
& \qquad \qquad\;\;\; = (-1)^{L+L'+L^{''}} \sum_{M,M',M^{''}} \begin{pmatrix}
L & L' & L^{''}\\
M & M' & M^{''}
\end{pmatrix}\nonumber \\
& \qquad \qquad \qquad  \times \int_{\hatk} Y_{L,M}(\hatk)  Y_{L',M'}(\hatk)  Y_{L^{''},M^{''}}(\hatk)  Y_{\ell,m}(\hatk) Y_{\ell',m'}(\hatk) \nonumber \\
& \qquad \qquad\;\;\;  = (-1)^{L+L'+L^{''}} \sum_{M,M',M^{''}}\;\sum_{\ell^{''},m^"}\sum_{\ell^{'''},m^{'''}} \begin{pmatrix}
L & L' & L^{''}\\
M & M' & M^{''}
\end{pmatrix} \nonumber \\
& \qquad \qquad \qquad  \times G_{L^{''}, \ell, \ell^{''}}^{M^{''},m,-m^{''}} G_{\ell',\ell^{''}, \ell^{'''}}^{m',m^{''},-m^{'''}} G_{L,L', \ell^{'''}}^{M,M',m^{'''}}.
\end{align}
Here, we have followed the same steps as in Eq. (\ref{eq:Fancy_H_def}), with the exception that, to reach the third equality, we needed to combine two spherical harmonics into a single one twice in order to obtain an integral over a product of three spherical harmonics. 

\section{Evaluation of $\hats$ Integrals}
\label{sec:S_evaluation}
In this section, we evaluate the $\hats$ integrals of the five configurations that contribute to the 1-loop covariance in $\S$\ref{sec:Cov_Compu}, following \cite{Cahn_Iso, Hou_Cov} by averaging the isotropic basis over rotations $R,\;R$, and $S$. Averaging over rotations $R$ and $R'$ enables to express the angular dependence in terms of a single isotropic basis function, while averaging over rotations $S$ enables the evaluation of the $\hats$ dependence as previously demonstrated in \cite{Cahn_Iso, Hou_Cov}.
\subsection{$\hats$ Integral for the Primary-Secondary Configuration}\label{sec:s_hat_PS_Evaluation}
We begin our analysis of these integrals with the Primary-Secondary configuration:
\begin{align}
&\int dR\;dR'\;dS \;\mathcal{P}_{L_1,L_{1s},L'_{1}}(R\hatr_1,-S\hats,-R'\hatr'_1)\mathcal{P}_{L_{30},L'_{3s},L'_{3}}(R\hatr_0,-S\hats,-R'\hatr'_3)\nonumber \\
&  \times \mathcal{P}_{L_{20},L'_{2s},L'_{2}}(R\hatr_0,-S\hats,-R'\hatr'_2)\mathcal{P}_{L_3,L_{3s},L'_{30}}(R\hatr_3,-S\hats,-R'\hatr'_0)\nonumber \\
&  \times \mathcal{P}_{L_2,L_{2s},L'_{20}}(R\hatr_2,-S\hats,-R'\hatr'_0). 
\end{align}
Expanding the isotropic functions into spherical harmonics and applying parity inversion, we find:
\begin{align}
&\int dR\;dR'\;dS \;(-1)^{L_{1}+L_2+L_3+L_{20}+L_{30}} \nonumber \\
&\qquad\times \sum_{{\rm All}\; M}\mathcal{C}^{L_{1},L_{1s},L'_{1}}_{M_{1},M_{1s},M'_{1}}\;\mathcal{C}^{L_{30},L'_{3s},L'_{3}}_{M_{30},M'_{3s},M'_{3}}\;\mathcal{C}^{L_{20},L'_{2s},L'_{2}}_{M_{20},M'_{2s},M'_{2}} \;\mathcal{C}^{L_{2},L_{2s},L'_{20}}_{M_{2},M_{2s},M'_{20}}\;\mathcal{C}^{L_{3},L_{3s},L'_{30}}_{M_{3},M_{3s},M'_{30}} \nonumber \\
&\qquad\times Y_{L_{20},M_{20}}(R\hatr_0) Y_{L_{30},M_{30}}(R\hatr_0)Y_{L_{1},M_{1}}(R\hatr_1)Y_{L_{2},M_{2}}(R\hatr_2)Y_{L_{3},M_{3}}(R\hatr_3)\nonumber \\
&\qquad\times Y_{L'_{2s},M'_{2s}}(S\hats) Y_{L'_{3s},M'_{3s}}(S\hats) Y_{L_{1s},M_{1s}}(S\hats)Y_{L_{2s},M_{2s}}(S\hats)Y_{L_{3s},M_{3s}}(S\hats) \nonumber \\
&\qquad\times Y_{L'_{2},M'_{2}}(R'\hatr'_2) Y_{L'_{3},M'_{3}}(R'\hatr'_3) Y_{L'_{1},M'_{1}}(R'\hatr'_1) Y_{L'_{20},M'_{20}}(R'\hatr'_0)Y_{L'_{30},M'_{30}}(R'\hatr'_0), 
\end{align}
where we define $\mathcal{C}$ as \cite{Hou_Cov}:
\begin{align}
&\mathcal{C}^{\Lambda}_M = \sqrt{2\ell_{12}+1}\times \cdots\times\sqrt{2\ell_{12\cdots n-2}+1}\nonumber\\
&\qquad \quad \times \sum_{m_{12},\cdots}(-1)^\kappa\tj{\ell_1}{\ell_2}{\ell_{12}}{m_1}{m_2}{-m_{12}} \tj{\ell_{12}}{\ell_3}{\ell_{123}}{m_{12}}{m_3}{-m_{123}} \cdots\nonumber \\
&\qquad \quad \times \tj{\ell_{12\cdots n-2}}{\ell_{n-1}}{\ell_n}{m_{12\cdots n-2}}{m_{n-1}}{m_n},
\end{align}
and $\Lambda = \{\ell_1,\ell_2,(\ell_{12}),\ell_3,(\ell_{123}),\cdots,\ell_n\}$ with the 'intermediate' angular momenta in brackets, $M = \{m_1,m_2,(m_{12}),m_3,(m_{123}),\cdots,m_n\}$, and $\kappa=\ell_{12}-m_{12}+\ell_{123}-m_{123}+\cdots+\ell_{12\ldots n-2}-m_{12... n-2}$.

We analyze the integrals one by one, starting with the integral over the rotational average of the $\hats$ vector, for which we find:
\begin{align}
&\int dS\;Y_{L'_{2s},M'_{2s}}(S\hats) Y_{L'_{3s},M'_{3s}}(S\hats) Y_{L_{1s},M_{1s}}(S\hats)Y_{L_{2s},M_{2s}}(S\hats)Y_{L_{3s},M_{3s}}(S\hats) \nonumber \\
& = \int d\hatn \;Y_{L'_{2s},M'_{2s}}(\hatn) Y_{L'_{3s},M'_{3s}}(\hatn) Y_{L_{1s},M_{1s}}(\hatn)Y_{L_{2s},M_{2s}}(\hatn)Y_{L_{3s},M_{3s}}(\hatn) \nonumber \\
& =\sum_{\ell,\ell'}\sum_{m,m'} G_{M'_{2s},M'_{3s},-m}^{L'_{2s},L'_{3s},\ell}G_{m, M_{1s},-m'}^{\ell,L_{1s},\ell'}G_{m', M_{2s},M_{3s}}^{\ell',L_{2s},L_{3s}},
\end{align}
where we have defined $\widehat{\mathbf{n}}\equiv S\hats$ in the first equality, and solved the resulting integral in the second equality in terms of the Gaunt integrals. 

Next, we evaluate the integral over the rotational average of the $\hatr'$ vectors, for which the solution is explicitly given in Eq. (2.31) of \cite{Cahn_Iso}. We state the result here:  
\begin{align}
&\int dR'\;Y_{L'_{2},M'_{2}}(R'\hatr'_2) Y_{L'_{3},M'_{3}}(R'\hatr'_3) Y_{L'_{1},M'_{1}}(R'\hatr'_1) Y_{L'_{20},M'_{20}}(R'\hatr'_0)Y_{L'_{30},M'_{30}}(R'\hatr'_0)\nonumber \\
&= \sum_{L'_0,M'_0}G_{M'_{20},M'_{30},-M'_0}^{L'_{20},L'_{30},L'_0}\sum_{\Lambda'} \mathcal{C}^{\Lambda'}_{M'}\PP_{\Lambda'}(\hatR'), 
\end{align}
where we have defined $\Lambda'=\left\{L'_{0},L'_{1},L'_{2},L'_{3}\right\}$ and likewise for $M'=\left\{M'_{0},M'_{1},M'_{2},M'_{3}\right\}$. We have combined the two spherical harmonics of the $\hatr'_0$ vectors into a single one in the second equality, and used Eq. (2.31) of \cite{Cahn_Iso} to evaluate the resulting integral with four spherical harmonics.

Lastly, we evaluate the integral over the rotational average of the $\hatr$ vectors. We follow the same procedure as for the $\hatr'$ evaluation:
\begin{align}
&\int dR'\;Y_{L_{20},M_{20}}(R\hatr_0) Y_{L_{30},M_{30}}(R\hatr_0)Y_{L_{1},M_{1}}(R\hatr_1)Y_{L_{2},M_{2}}(R\hatr_2)Y_{L_{3},M_{3}}(R\hatr_3)\nonumber \\
&= \sum_{L_0,M_0}G_{M_{20},M_{30},-M_0}^{L_{20},L_{30},L_0}\sum_{\Lambda} \mathcal{C}^{\Lambda}_{M}\PP_{\Lambda}(\hatR), 
\end{align}
where we have defined $\Lambda=\left\{L_{0},L_{1},L_{2},L_{3}\right\}$ and $M=\left\{M_{0},M_{1},M_{2},M_{3}\right\}$. 

Therefore, the result of the $\hats$ integral for the primary-secondary configuration can be written as:
\begin{align}\label{eq:s_hat_PS}
&\int dR\;dR'\;dS \;\PP_{L_{20},L'_{2s},L'_{2}}(R\hatr_0,-S\hats,-R'\hatr'_2) \nonumber \\
&\times\PP_{L_{30},L'_{3s},L'_{3}}(R\hatr_0,-S\hats,-R'\hatr'_3)\PP_{L_{1},L_{1s},L'_{1}}(R\hatr_1,-S\hats,-R'\hatr'_1)\nonumber \\
&\times\PP_{L_{2},L_{2s},L'_{20}}(R\hatr_2,-S\hats,-R'\hatr'_0)\PP_{L_{3},L_{3s},L'_{30}}(R\hatr_3,-S\hats,-R'\hatr'_0)\nonumber \\
& = \sum_{\Lambda,\Lambda_s,\Lambda'}\overline{\mathcal{Q}}_{(\rm P-S)}^{\Lambda,\Lambda_s,\Lambda'} \;\PP_{\Lambda} (\hatR)\PP_{\Lambda'}(\hatR'), 
\end{align}
where we have defined $\Lambda_s =\left\{L'_{2s},L'_{3s},L_{1s},L_{2s},L_{3s}\right\}$ and $\overline{\mathcal{Q}}$ as:
\begin{align}\label{eq:Q_bar_def_SS}
&\overline{\mathcal{Q}}_{(\rm P-S)}^{\Lambda,\Lambda_s,\Lambda'} \equiv(-1)^{L_{1}+L_2+L_3+L_{20}+L_{30}} \nonumber \\
& \qquad \quad  \qquad \;\;\times\sum_{\rm All\; M}\mathcal{C}^{L_{20},L'_{2s},L'_{2}}_{M_{20},M_{2s},M'_{2}}\;\mathcal{C}^{L_{30},L'_{3s},L'_{3}}_{M_{30},M_{3s},M'_{3}}\;\mathcal{C}^{L_{1},L_{1s},L'_{1}}_{M_{1},M_{1s},M'_{1}} \;\mathcal{C}^{L_{2},L_{2s},L'_{20}}_{M_{2},M_{2s},M'_{20}}\;\mathcal{C}^{L_{3},L_{3s},L'_{30}}_{M_{3},M_{3s},M'_{30}}\nonumber \\
& \qquad \quad  \qquad \;\;\times \sum_{L_0,L'_0}\sum_{M_0,M'_0}G_{M_{20},M_{30},-M_0}^{L_{20},L_{30},L_0} \mathcal{C}^{\Lambda}_{M}\;G_{M'_{20},M'_{30},-M'_0}^{L'_{20},L'_{30},L'_0} \mathcal{C}^{\Lambda'}_{M'}\nonumber \\
& \qquad \quad  \qquad \;\; \times \sum_{\ell,\ell'}\sum_{m,m'} G_{M_{2s},M_{3s},-m}^{L'_{2s},L'_{3s},\ell}G_{m, M_{1s},-m'}^{\ell,L_{1s},\ell'}G_{m', M_{2s},M_{3s}}^{\ell',L_{2s},L_{3s}}.
\end{align}

\subsection{$\hats$ Integral for the Half-Half Configuration}\label{sec:s_hat_HH_Evaluation}
We continue the analysis of the $\hats$ integrals by evaluating the half-half configuration:
\begin{align}
&\int dR\;dR'\;dS \;\mathcal{P}_{L_{q0},L’_{qs},L’_{q’_0}}(-R\hatr_0,S\hats,R'\hatr’_0) \PP_{L_{30},L’_{3s},L’_{3},}(-R\hatr_0,S\hats,R'\hatr’_3)\nonumber \\
&\quad \times \PP_{L_1,L_{1s},L’_1}(R\hatr_1,-S\hats,-R'\hatr’_1)\PP_{L_2,L_{2s},L’_2}(R\hatr_2,-S\hats,-R'\hatr’_2)  \nonumber \\
&\quad \times \mathcal{P}_{L_3,L_{3s},L'_{30}}(R\hatr_3,-S\hats,-R'\hatr'_0). 
\end{align}
The result of evaluating the above integral will be analogous to the primary-secondary configuration, the only difference being that $\Lambda_s = \left\{L'_{qs},L'_{3s},L_{1s},L_{2s},L_{3s}\right\}$, and the $\mathcal{Q}$ constant is:
\begin{align}\label{eq:Q_bar_def_HH}
&\overline{\mathcal{Q}}_{(\rm H-H)}^{\Lambda,\Lambda_s,\Lambda'} \equiv(-1)^{L_{1}+L_2+L_3+L'_{qs}+L'_{q'_0} + L'_{3s}+L'_3} \sum_{\rm All\; M}\mathcal{C}^{L_{q0},L'_{qs},L'_{q'_0}}_{M_{q0},M'_{qs},M'_{q'_0}}\;\mathcal{C}^{L_{30},L'_{3s},L'_{3}}_{M_{30},M'_{3s},M'_{3}}\;\mathcal{C}^{L_{1},L_{1s},L'_{1}}_{M_{1},M_{1s},M'_{1}} \nonumber \\
& \qquad \qquad  \;\;\times \mathcal{C}^{L_{2},L_{2s},L'_{2}}_{M_{2},M_{2s},M'_{2}}\mathcal{C}^{L_{3},L_{3s},L'_{30}}_{M_{3},M_{3s},M'_{30}}\sum_{L_0,L'_0}\sum_{M_0,M'_0}G_{M_{q0},M_{30},-M_0}^{L_{q0},L_{30},L_0} \mathcal{C}^{\Lambda}_{M}\;G_{M'_{q0},M'_{30},-M'_0}^{L'_{q0},L'_{30},L'_0} \mathcal{C}^{\Lambda'}_{M'}\nonumber \\
& \qquad \qquad \;\; \times \sum_{\ell,\ell'}\sum_{m,m'} G_{M'_{qs},M'_{3s},-m}^{L'_{qs},L'_{3s},\ell}G_{m, M_{1s},-m'}^{\ell,L_{1s},\ell'}G_{m', M_{2s},M_{3s}}^{\ell',L_{2s},L_{3s}}.
\end{align}

\subsection{$\hats$ Integral for the Primary-Primary Configuration}\label{sec:s_hat_PP_Evaluation}
Next, we evaluate the primary-primary configuration $\hats$ integral, given by:
\begin{align}
&\int dR\;dR'\;dS \;\PP_{L_{01},L_{1qs},L'_{01}}(R\hatr_0,-S\hats,-R'\hatr'_0)\PP_{L_{02},L_{2qs},L'_{02}}(R\hatr_0,-S\hats,-R'\hatr'_0) \nonumber \\
&\qquad \times \PP_{L_{1},L_{1s},L'_{1}}(R\hatr_1,-S\hats,-R'\hatr'_1) \PP_{L_{2},L_{2s},L'_{2}}(R\hatr_2,-S\hats,-R'\hatr'_2)\nonumber \\
&\qquad \times \PP_{L_{3},L_{3s},L'_{3}}(R\hatr_3,-S\hats,-R'\hatr'_3). 
\end{align}
Evaluating these integrals, like the half-half configuration, follows the same steps as shown in the primary-secondary configuration. Once again, $\Lambda_s$ changes to $\Lambda_s = \left\{L_{1qs},L'_{2qs},L_{1s},L_{2s},L_{3s}\right\}$, and the $\mathcal{Q}$ constant is:
\begin{align}\label{eq:Q_bar_def_PP}
&\overline{\mathcal{Q}}_{(\rm P-P)}^{\Lambda,\Lambda_s,\Lambda'} \equiv(-1)^{L_{1}+L_2+L_3+L_{01}+L_{02}} \sum_{\rm All\; M}\mathcal{C}^{L_{01},L'_{1qs},L'_{01}}_{M_{01},M_{1qs},M'_{01}}\mathcal{C}^{L_{02},L_{2qs},L'_{02}}_{M_{02},M_{2qs},M'_{02}} \nonumber \\
& \qquad \qquad  \;\;\times \mathcal{C}^{L_{1},L_{1s},L'_{1}}_{M_{1},M_{1s},M'_{1}} \mathcal{C}^{L_{2},L_{2s},L'_{2}}_{M_{2},M_{2s},M'_{2}}\mathcal{C}^{L_{3},L_{3s},L'_{3}}_{M_{3},M_{3s},M'_{3}}\sum_{L_{0},L'_{0}}\sum_{M_0,M'_0}G_{M_{01},M_{02},-M_0}^{L_{01},L_{02},L_0} \mathcal{C}^{\Lambda}_{M}\;G_{M'_{01},M'_{02},-M'_0}^{L'_{01},L'_{02},L'_0} \mathcal{C}^{\Lambda'}_{M'}\nonumber \\
& \qquad \qquad \;\; \times \sum_{\ell,\ell'}\sum_{m,m'} G_{M_{1qs},M_{2qs},-m}^{L_{1qs},L'_{2qs},\ell}G_{m, M_{1s},-m'}^{\ell,L_{1s},\ell'}G_{m', M_{2s},M_{3s}}^{\ell',L_{2s},L_{3s}}. 
\end{align}

\subsection{$\hats$ Integral for the PC-Primary-Secondary Configuration}\label{sec:s_hat_PC-PS_Evaluation}
We continue the analysis of the $\hats$ integrals by evaluating the PC-primary-secondary configurations:
\begin{align}
&\int dR\;dR'\;dS \;Y_{L_{10},M_{10}}(R\hatr_0)Y_{L_{11},M_{11}}(R\hatr_1) Y_{L_{20},M_{20}}(R\hatr_0)Y_{L_{22},M_{22}}(R\hatr_2) Y_{L'_{10},M'_{10}}(R'\hatr'_0)\nonumber \\
& \times Y_{L'_{11},M'_{11}}(R'\hatr'_1) Y_{L'_{20},M'_{20}}(R'\hatr'_0)Y_{L'_{22},M'_{22}}(R'\hatr'_2) \PP_{L_3,L_{3s},L'_{3}}(R\hatr_3,-S\hats,-R'\hatr'_3). 
\end{align}
Expressing the 3-argument isotropic basis function in terms of the spherical harmonics shows that the result of the $dS$ integral gives $L_{3s}=0$; through the 3-$j$ symbol from the expansion, we then obtain $L_3 = L'_3$. The primed integrals can be performed to obtain:
\begin{align}
&(-1)^{L_3}\int dR'\;Y_{L'_{10},M'_{10}}(R'\hatr'_0)Y_{L'_{20},M'_{20}}(R'\hatr'_0) Y_{L'_{11},M'_{11}}(R'\hatr'_1)Y_{L'_{22},M'_{22}}(R'\hatr'_2)Y_{L_3,M_3}(R'\hatr'_3)\nonumber \\
&= (-1)^{L_3}\sum_{L'_0,M'_0} G^{L'_{10},L'_{20},L'_0}_{M'_{10},M'_{20},-M'_0} \int dR\; Y_{L'_0,M'_0}(R\hatr_0)Y_{L'_{11},M'_{11}}(R'\hatr'_1)Y_{L'_{22},M'_{22}}(R'\hatr'_2)Y_{L_3,M_3}(R'\hatr'_3)\nonumber \\
&=(-1)^{L_3}\sum_{\Lambda'}\mathcal{C}^{\Lambda'}_{M'}\PP_{\Lambda'}(\hatr'_0,\hatr'_1,\hatr'_2,\hatr'_3) \sum_{L'_0,M'_0}G^{L'_{10},L'_{20},L'_0}_{M'_{10},M'_{20},-M'_0},
\end{align}
where we have defined $\Lambda' \equiv \left\{L'_0,L'_{11},L'_{22},L_3\right\}$ and $M' \equiv \left\{M'_0,M'_{11},L'_{22},M_3\right\}$. Following the same steps, we evaluate the unprimed integrals and find:
\begin{align}
&\int dR\;Y_{L_{10},M_{10}}(R\hatr_0)Y_{L_{20},M_{20}}(R\hatr_0) Y_{L_{11},M_{11}}(R\hatr_1)Y_{L_{22},M_{22}}(R\hatr_2)Y_{L_3,M_3}(R\hatr_3)\nonumber \\
&=\sum_{\Lambda}\mathcal{C}^{\Lambda}_{M}\PP_{\Lambda}(\hatr_0,\hatr_1,\hatr_2,\hatr_3) \sum_{L_0,M_0}G^{L_{10},L_{20},L_0}_{M_{10},M_{20},-M_0},
\end{align}
where we have defined $\Lambda \equiv \left\{L_0,L_{10},L_{20},L_3\right\}$ and $M \equiv \left\{M_0,M_{11},L_{22},M_3\right\}$. Therefore, the result of evaluating the $\hats$ by means of averaging over all rotations results in:
\begin{align}\label{eq:s_hat_PC-PS_result}
&\int dR\;dR'\;dS \;Y_{L_{10},M_{10}}(R\hatr_0)Y_{L_{11},M_{11}}(R\hatr_1) Y_{L_{20},M_{20}}(R\hatr_0)Y_{L_{22},M_{22}}(R\hatr_2) Y_{L'_{10},M'_{10}}(R'\hatr'_0)\nonumber \\
& \times Y_{L'_{11},M'_{11}}(R'\hatr'_1) Y_{L'_{20},M'_{20}}(R'\hatr'_0)Y_{L'_{22},M'_{22}}(R'\hatr'_2) \PP_{L_3,L_{3s},L'_{3}}(R\hatr_3,-S\hats,-R'\hatr'_3)\nonumber \\
&= \sum_{\Lambda,\Lambda'} \;\overline{\mathcal{Q}}_{(\rm PC-P-S)}^{\Lambda,\Lambda',M,M'}\PP_{\Lambda}(\hatR)\PP_{\Lambda'}(\hatR'),
\end{align}
where we have defined:
\begin{align}\label{eq:Q_bar_def_PC-PS}
&\overline{\mathcal{Q}}_{(\rm PC-P-S)}^{\Lambda,\Lambda',M,M'} \equiv \sum_{M_3}\mathcal{C}^{\Lambda}_{M}\;\mathcal{C}^{\Lambda'}_{M'}(-1)^{L_3}D_{L_3}\sum_{L_0,M_0}G^{L_{10},L_{20},L_0}_{M_{10},M_{20},-M_0}\sum_{L'_0,M'_0}G^{L'_{10},L'_{20},L'_0}_{M'_{10},M'_{20},-M'_0}.
\end{align}

\subsection{$\hats$ Integral for the PC-Half-Half Configuration}\label{sec:s_hat_PC-HH_Evaluation}
We conclude our analysis of the $\hats$ integrals by evaluating the PC-half-half configurations:
\begin{align}
&\int dR\;dR'\;dS \;Y_{L_{10},M_{10}}(R\hatr_0)Y_{L_{11},M_{11}}(R\hatr_1)Y_{L'_{10},M'_{10}}(R'\hatr'_0)Y_{L'_{11},M'_{11}}(R'\hatr'_1) \nonumber \\ 
& \times \PP_{L_{q20},L_{q2s},L'_{q20}}(R\hatr_0,-S\hats,-R'\hatr'_0) \PP_{L_2,L_{2s},L'_{2}}(R\hatr_2,-S\hats,-R'\hatr'_2)\nonumber \\ 
& \times \PP_{L_3,L_{3s},L'_{3}}(R\hatr_3,-S\hats,-R'\hatr'_3). 
\end{align}
Expressing the 3-argument isotropic basis function in terms of the spherical harmonics we find the $dS$ integral as:
\begin{align}
\int dS\;Y_{L_{q2s},M_{q2s}}(\hats)Y_{L_{2s},M_{2s}}(\hats)Y_{L_{3s},M_{3s}}(\hats) = G_{M_{q2s},M_{2s},M_{3s}}^{L_{q2s},L_{2s},L_{3s}}.
\end{align}
Then, the primed integrals can be evaluated to obtain:
\begin{align}
&\int dR'\;Y_{L'_{10},M'_{10}}(R'\hatr'_0)Y_{L'_{q20},M'_{q20}}(R'\hatr'_0) Y_{L'_{11},M'_{11}}(R'\hatr'_1)Y_{L'_{2},M'_{2}}(R'\hatr'_2)Y_{L'_3,M'_3}(R'\hatr'_3)\nonumber \\
&= \sum_{L'_0,M'_0} G^{L'_{10},L'_{q20},L'_0}_{M'_{10},M'_{q20},-M'_0} \int dR\; Y_{L'_0,M'_0}(R\hatr_0)Y_{L'_{11},M'_{11}}(R'\hatr'_1)Y_{L'_{2},M'_{2}}(R'\hatr'_2)Y_{L'_3,M'_3}(R'\hatr'_3)\nonumber \\
&=\sum_{\Lambda'}\mathcal{C}^{\Lambda'}_{M'}\PP_{\Lambda'}(\hatr'_0,\hatr'_1,\hatr'_2,\hatr'_3) \sum_{L'_0,M'_0}G^{L'_{10},L'_{q20},L'_0}_{M'_{10},M'_{q20},-M'_0},
\end{align}
where we have defined $\Lambda' \equiv \left\{L'_0,L'_{11},L'_{2},L'_3\right\}$ and $M' \equiv \left\{M'_0,M'_{11},M'_{2},M'_3\right\}$. Following the same steps, we evaluate the unprimed integrals and find:
\begin{align}
&\int dR\;Y_{L_{10},M_{10}}(R\hatr_0)Y_{L_{q20},M_{2q0}}(R\hatr_0) Y_{L_{11},M_{11}}(R\hatr_1)Y_{L_{2},M_{2}}(R\hatr_2)Y_{L_3,M_3}(R\hatr_3)\nonumber \\
&=\sum_{\Lambda}\mathcal{C}^{\Lambda}_{M}\PP_{\Lambda}(\hatr_0,\hatr_1,\hatr_2,\hatr_3) \sum_{L_0,M_0}G^{L_{10},L_{q20},L_0}_{M_{10},M_{q20},-M_0},
\end{align}
where we have defined $\Lambda \equiv \left\{L_0,L_{11},L_{2},L_3\right\}$ and $M \equiv \left\{M_0,M_{11},M_{2},M_3\right\}$. Therefore, the result of evaluating the $\hats$ by means of averaging over all rotations results in:
\begin{align}\label{eq:s_hat_PC-HH_result}
&\int dR\;dR'\;dS \;Y_{L_{10},M_{10}}(R\hatr_0)Y_{L_{11},M_{11}}(R\hatr_1)Y_{L'_{10},M'_{10}}(R'\hatr'_0)Y_{L'_{11},M'_{11}}(R'\hatr'_1) \nonumber \\ 
& \times \PP_{L_{q20},L_{q2s},L'_{q20}}(R\hatr_0,-S\hats,-R'\hatr'_0) \PP_{L_2,L_{2s},L'_{2}}(R\hatr_2,-S\hats,-R'\hatr'_2)\nonumber \\ 
& \times \PP_{L_3,L_{3s},L'_{3}}(R\hatr_3,-S\hats,-R'\hatr'_3)\nonumber \\
&= \sum_{\Lambda,\Lambda_s,\Lambda'} \;\overline{\mathcal{Q}}_{(\rm PC-H-H)}^{\Lambda,\Lambda_s,\Lambda'}\PP_{\Lambda}(\hatR)\PP_{\Lambda'}(\hatR'),
\end{align}
where $\Lambda_s \equiv \left\{L_{q2s},L_{2s},L_{3s}\right\}$ and $M_s \equiv \left\{M_{q2s},M_{2s},M_{3s}\right\}$, and we have defined:
\begin{align}\label{eq:Q_bar_def_PC-HH}
&\overline{\mathcal{Q}}_{(\rm PC-H-H)}^{\Lambda,\Lambda_s,\Lambda', \bar{M}} \equiv (-1)^{L_{q20}+L_2+L_3}\sum_{M_{q20},M_{q2s},M'_{q20}}\sum_{M_2,M_{2s},M'_2}\sum_{M_3,M_{3s},M'_3}\mathcal{C}^{\Lambda}_{M}\;\mathcal{C}^{\Lambda'}_{M'}\mathcal{C}^{L_{q20},L_{q2s},L'_{q20}}_{M_{q20},M_{q2s},M'_{q20}}\nonumber \\
&\qquad\qquad \qquad \times \mathcal{C}^{L_{2},L_{2s},L'_{2}}_{M_{2},M_{2s},M'_{2}}\mathcal{C}^{L_{3},L_{3s},L'_{3}}_{M_{3},M_{3s},M'_{3}}\sum_{L_0,M_0}G^{L_{10},L_{q20},L_0}_{M_{10},M_{q20},-M_0}\sum_{L'_0,M'_0}G^{L'_{10},L'_{q20},L'_0}_{M'_{10},M'_{q20},-M'_0}.
\end{align}
Above we have defined $\bar{M}=\{M_{10},M_{11},M'_{10},M'_{11}\}$ since these magnetic quantum numbers are not summed until their final evaluation in Eq. (\ref{eq:PC-HH_res}). 

\section{Forming the Connected 4PCF Covariance}\label{sec:Form_4PCF_Cov}
We show how to form the connected 4PCF covariance terms entering at the 1-loop order. Following \cite{Philcox_4PCF_measurement}, every integral over correlation functions scales as $r_{\text{c}}^3/V$, where $r_{\text{c}} \approx 100$ Mpc/$h$ is the correlation length and $V$ is the volume of the survey. The survey size tends to be much larger than the correlation length, hence for every integral involving a correlation function, we have a suppression of the order of $\left(r_{\text{c}}^3/V\right)^{\alpha}$, where $\alpha = \left\{1,2,3,\cdots\right\}$ indicates the number of integrals over correlation functions we have. 

Therefore, we compute the connected 4PCF covariance with only the terms that are of the same order in $r_{\text{c}}^3/V$. We first define the components that give rise to the connected 4PCF, starting with the full 4PCF:
\begin{align}
&\hat\zeta^{(\rm full)}(\vecr_0,\vecr_1,\vecr_2,\vecr_3) = \frac{1}{V}\int d^3\vecx\,\delta(\vecx+\vecr_0)\delta(\vecx+\vecr_1)\delta(\vecx+\vecr_2)\delta(\vecx+\vecr_3),
\end{align}
following the convention of the main text where $\vecr_0 = 0$, but we keep it for the symmetry of the problem. Likewise, the disconnected piece gives:
\begin{align}
&\hat\zeta^{(\rm disc.)}(\vecr_0, \vecr_1,\vecr_2,\vecr_3) = \frac{1}{V}\int d^3\vecx_1\,\delta(\vecx_1+\vecr_0)\delta(\vecx_1+\vecr_1)\;\frac{1}{V}\int d^3\vecx_2 \;\delta(\vecx_2+\vecr_2)\delta(\vecx_2+\vecr_3). 
\end{align}
Given the above definitions, to compute the connected 4PCF covariance we must then obtain:
\begin{align}\label{eq:Connected_Cov_definition}
&\mathrm{Cov}_{4,\text{1L}}^{[2],( c, c)} = \mathrm{Cov_{4,\text{1L}}}^{[2],(\rm full, full)} - \mathrm{Cov}_{4,\text{1L}}^{[2],(\rm full, disc)} - \mathrm{Cov}_{4,\text{1L}}^{[2],(\rm disc, full)} + \mathrm{Cov}_{4,\text{1L}}^{[2],(\rm disc, disc)}. 
\end{align}
We show how to calculate $\mathrm{Cov}^{(\rm full, full)}$ with the second-order density contrast, but computing the other terms follows the same procedure:
\begin{align}
&\mathrm{Cov}_{4,\text{1L}}^{[2],(\rm full, full)} = \left\langle \frac{1}{V}\int d^3\vecx\,\delta^{(2)}(\vecx+\vecr_0)\dlin(\vecx+\vecr_1)\dlin(\vecx+\vecr_2)\dlin(\vecx+\vecr_3)\right.\nonumber \\
& \qquad \qquad \qquad \left. \times \frac{1}{V}\int d^3\vecx'\,\delta^{(2)}(\vecx'+\vecr'_0)\dlin(\vecx'+\vecr'_1)\dlin(\vecx'+\vecr'_2)\dlin(\vecx'+\vecr'_3)\right\rangle.
\end{align}
We expand the second-order densities in terms of two linear densities in real space\footnote{We set the second-order kernel to unity to demonstrate how to construct the connected 4PCF covariance. In the main text, we evaluate the covariance with the second-order kernel incorporated.} and apply Wick's theorem to obtain:
\begin{align}\label{eq:Cov_FullxFull_term}
&\mathrm{Cov}_{4,\text{1L}}^{[2],(\rm full, full)} \supset \frac{1}{V^2}\int d^3\vecq_1\;d^3\vecq_2\int d^3\vecq'_1\;d^3\vecq'_2\int d^3\vecx\int d^3\vecx'\left[\left\{ \left\langle \dlin(\vecx+\vecr_0+\vecq_1) \dlin (\vecx'+\vecr'_0+\vecq'_1) \right\rangle \right. \right. \nonumber \\
& \qquad \qquad \qquad\left.\left. \times \left\langle \dlin(\vecx+\vecr_0+\vecq_2) \dlin (\vecx'+\vecr'_0+\vecq'_2) \right\rangle \left\langle \dlin(\vecx+\vecr_1) \dlin (\vecx'+\vecr'_1) \right\rangle \right. \right. \nonumber \\
& \qquad \qquad \qquad\left. \left. \times\left\langle \dlin(\vecx+\vecr_2) \dlin (\vecx'+\vecr'_2) \right\rangle \left\langle \dlin(\vecx+\vecr_3) \dlin (\vecx'+\vecr'_3) \right\rangle + \text{137 perms.}\footnotemark[9] \right\} \right. \nonumber \\
& \qquad \qquad \qquad  + \left\{ \left. \left\langle \dlin(\vecx+\vecr_0+\vecq_1) \dlin (\vecx+\vecr_1) \right\rangle \left\langle \dlin(\vecx+\vecr_0+\vecq_2) \dlin (\vecx'+\vecr'_0+\vecq'_2) \right\rangle\right. \right. \nonumber \\ 
& \qquad \qquad \qquad\left. \left. \times \left\langle \dlin(\vecx'+\vecq'_1+\vecr'_0) \dlin (\vecx'+\vecr'_1) \right\rangle \left\langle \dlin(\vecx+\vecr_2) \dlin (\vecx'+\vecr'_2) \right\rangle\right.\right. \nonumber \\
& \qquad \qquad \qquad\left. \left. \times \left\langle \dlin(\vecx+\vecr_3) \dlin (\vecx'+\vecr'_3) \right\rangle + \text{419 perms.}\footnotemark[10] \right\} \right] \nonumber \\
& \qquad \qquad \;\; = \frac{1}{V}\int d^3\vecq_1\;d^3\vecq_2\int d^3\vecq'_1\;d^3\vecq'_2\int d^3\vecs \nonumber \\
& \qquad \qquad \qquad \times \left[ \left\{\xi_W( \vecr_0-\vecs-\vecr'_0+\vecq_1-\vecq'_1)\xi_W( \vecr_0-\vecs-\vecr'_0+\vecq_2-\vecq'_2)\right.\right. \nonumber \\
& \qquad \qquad \qquad\left.\left. \times\xi( \vecr_1-\vecs-\vecr'_1)\xi( \vecr_2-\vecs-\vecr'_2)\xi( \vecr_3-\vecs-\vecr'_3)+ \text{119 perms.}\right\} \right. \nonumber \\
& \qquad \qquad \qquad  + \left. \left\{ \xi_W(\vecr_0-\vecr_1+\vecq_1)\xi_W(\vecr_0-\vecs-\vecr'_0+\vecq_2-\vecq'_2)\xi_W(\vecr'_0-\vecr'_1+\vecq'_1)\right.\right. \nonumber \\
& \qquad \qquad \qquad\left. \left.\times\xi( \vecr_2-\vecs-\vecr'_2)\xi( \vecr_3-\vecs-\vecr'_3)+ \text{419 perms.}\right\}\right],
\end{align}
\footnotetext[9]{The number of permutations has been obtained by adding the permutations counted in the main text for the fully-coupled terms.} \footnotetext[10]{Finding the number of permutations is done in three steps. i) Choose one of the second-order densities in either of the two tetrahedra, then pair this second-order density with any of the densities from the same tetrahedron; one has 4 possibilities. ii) Pair the remaining second-order density with any density in any of the two tetrahedrons; one has 7 possibilities. iii) Repeat with the opposite tetrahedron; one has 15 possibilities. Total number of permutations = 4 $\times$ 7$\times$ 15 = 420 perms.}\\
\noindent where we express the correlation function derived from the second-order densities as $\xi_W$ to indicate that this must actually be evaluated with their respective second-order kernels as shown in the main text. The correlation functions in the last two lines arise from contracting density contrasts within the same tetrahedron; we term these structures partially-connected correlations. All the above permutations are of the order $\mathcal{O}(r_c^3/V)$. 

According to Eq. (\ref{eq:Connected_Cov_definition}) all the terms with the same order, $\mathcal{O}(r_c^3/V)$, will cancel each other; terms that are suppressed in the full $\times$ disconnected, disconnected $\times$ full, and disconnected $\times$ disconnected pieces will not cancel the full $\times$ full piece. This means we must check the full $\times$ disconnected, disconnected $\times$ full, and disconnected $\times$ disconnected pieces carefully to determine which terms enter the connected 4PCF covariance at the 1-loop level. 

We find that only the two partially-connected configurations depicted in Figure~\ref{fig:2nd_pc_terms} enter the covariance. Below, we show an example of one of the partially-connected permutations that is being suppressed in the full $\times$ disconnected, disconnected $\times$ full, and disconnected $\times$ disconnected pieces, hence entering the covariance:
\begin{align}
&\mathrm{Cov}_{4,\text{1L}}^{[2],(\rm full, disc.)} \supset \frac{1}{V^3}\int d^3\vecq_1\;d^3\vecq_2\int d^3\vecq'_1\;d^3\vecq'_2\int d^3\vecx\int d^3\vecx_1'\int d^3\vecx_2' \nonumber \\ 
& \qquad \qquad \qquad  \times \left[\left\langle \dlin(\vecx+\vecr_0+\vecq_1) \dlin (\vecx+\vecr_1) \right\rangle \left\langle \dlin(\vecx+\vecr_0+\vecq_2) \dlin (\vecx_1'+\vecr'_0+\vecq'_2) \right\rangle \right. \nonumber \\
& \qquad \qquad \qquad \left. \left\langle \dlin(\vecx_1'+\vecq'_1+\vecr'_0) \dlin (\vecx_1'+\vecr'_1) \right\rangle \left\langle \dlin(\vecx+\vecr_2) \dlin (\vecx_2'+\vecr'_2) \right\rangle \right. \nonumber \\
& \qquad \qquad \qquad \left. \times \left\langle \dlin(\vecx+\vecr_3) \dlin (\vecx_2'+\vecr'_3) \right\rangle + \text{71 perms.}\footnotemark  \right] \nonumber \\
& \qquad \qquad \;\; = \frac{1}{V^2}\int d^3\vecq_1\;d^3\vecq_2\int d^3\vecq'_1\;d^3\vecq'_2\int d^3\vecs \int d^3\vecs' \nonumber \\
& \qquad \qquad \qquad \times\left[ \xi_W(\vecr_0-\vecr_1+\vecq_1)\xi_W(\vecr_0-\vecs-\vecr'_0+\vecq_2-\vecq'_2)\right. \nonumber \\
& \qquad \qquad \qquad \left. \times\xi_W(\vecr'_0-\vecr'_1+\vecq'_1)\xi( \vecr_2-\vecs'-\vecr'_2)\xi( \vecr_3-\vecs'-\vecr'_3)+ \text{71 perms.}\right],
\end{align}
\footnotetext{The number of permutations has been obtained by adding the permutations counted in the main text for the partially-coupled terms.}where we defined $\vecs = \vecx - \vecx'_1$ and $\vecs'=\vecx-\vecx_2'$. The above result gets suppressed by $r_{\text{c}}^6/V^2$. 

Calculating the disconnected $\times$ full term of the covariance follows the same procedure, and results in the same number of permutations being suppressed:  
\begin{align}
&\mathrm{Cov}_{4,\text{1L}}^{[2],(\rm disc., disc.)} \supset \frac{1}{V^4}\int d^3\vecq_1\;d^3\vecq_2\int d^3\vecq'_1\;d^3\vecq'_2\int d^3\vecx_1\int d^3\vecx_2\int d^3\vecx_1'\int d^3\vecx_2' \nonumber \\ 
& \qquad \qquad \qquad  \times\left[\left\langle \dlin(\vecx_1+\vecr_0+\vecq_1) \dlin (\vecx_1+\vecr_1) \right\rangle \left\langle \dlin(\vecx_1+\vecr_0+\vecq_2) \dlin (\vecx_1'+\vecr'_0+\vecq'_2) \right\rangle \right. \nonumber \\
& \qquad \qquad \qquad \left. \times \left\langle \dlin(\vecx_1'+\vecq'_1+\vecr'_0) \dlin (\vecx_1'+\vecr'_1) \right\rangle \left\langle \dlin(\vecx_2+\vecr_2) \dlin (\vecx_2'+\vecr'_2) \right\rangle\right. \nonumber \\
& \qquad \qquad \qquad \left. \times \left\langle \dlin(\vecx_2+\vecr_3) \dlin (\vecx_2'+\vecr'_3) \right\rangle + \text{71 perms.}  \right] \nonumber \\
& \qquad \qquad \;\; = \frac{1}{V^2}\int d^3\vecq_1\;d^3\vecq_2\int d^3\vecq'_1\;d^3\vecq'_2\int d^3\vecs_1 \int d^3\vecs_2' \nonumber \\
& \qquad \qquad \qquad \times\left[ \xi_W(\vecr_0-\vecr_1+\vecq_1)\xi_W(\vecr_0-\vecs_1-\vecr'_0+\vecq_2-\vecq'_2) \right. \nonumber \\
& \qquad \qquad \qquad \times \left.\xi_W(\vecr'_0-\vecr'_1+\vecq'_1)\xi( \vecr_2-\vecs_2'-\vecr'_2)\xi( \vecr_3-\vecs_2'-\vecr'_3)+ \text{71 perms.}\right],
\end{align}
where we defined $\vecs_1 = \vecx_1 - \vecx'_1$ and $\vecs_2'=\vecx_2-\vecx_2'$. This term is also suppressed at $\mathcal{O}(r_{\text{c}}^3/V)$. As anticipated, all of the partially-connected terms at order of $\mathcal{O}(r^3_{\text{c}}/V)$ will cancel out in Eq. (\ref{eq:Connected_Cov_definition}), while those terms that are being suppressed will not cancel out and hence need to be accounted for when computing the connected 4PCF covariance. Below we show an example of the terms that one needs to account for:
\begin{align}
 &\mathrm{Cov}_{4,\text{1L}}^{[2],({\text{c}},{\text{c}})} = \frac{1}{V}\int d^3\vecq_1\;d^3\vecq_2\int d^3\vecq'_1\;d^3\vecq'_2\int d^3\vecs \left[ \left\{\xi_W( \vecr_0-\vecs-\vecr'_0+\vecq_1-\vecq'_1)\xi_W( \vecr_0-\vecs-\vecr'_0+\vecq_2-\vecq'_2)\right.\right. \nonumber \\
& \qquad \qquad \qquad\left.\left. \times\xi( \vecr_1-\vecs-\vecr'_1)\xi( \vecr_2-\vecs-\vecr'_2)\xi( \vecr_3-\vecs-\vecr'_3)+ \text{119 perms.}\right\} \right. \nonumber \\
& \qquad \qquad \qquad  + \left. \left\{ \xi_W(\vecr_0-\vecr_1+\vecq_1)\xi_W(\vecr_0-\vecs-\vecr'_0+\vecq_2-\vecq'_2)\xi_W(\vecr'_0-\vecr'_1+\vecq'_1)\right.\right. \nonumber \\
& \qquad \qquad \qquad\left. \left.\times\xi( \vecr_2-\vecs-\vecr'_2)\xi( \vecr_3-\vecs-\vecr'_3)+ \text{71 perms.}\right\}\right]. 
\end{align}

\section{Next-to-Leading-Order Covariance Correction Scales}
\label{sec:Correction_Scale}
The NLO contribution to the 4PCF covariance matrix is suppressed relative to the Gaussian contribution by two additional powers of the linear density contrast,
\begin{equation}
\label{eq:NLO_scaling_delta}
\mathrm{Cov}_{\mathrm{NLO}} \sim \mathcal{O}\!\left(\delta_{\rm lin}^2\right)\times\mathrm{Cov}_{\mathrm{G}} .
\end{equation}
This scaling follows from standard perturbation theory, where the Gaussian covariance is sourced by disconnected products of two-point functions, while the NLO covariance receives contributions from connected diagrams involving higher-order mode couplings.

To make this statement quantitative, we estimate the variance of the linear density field smoothed on a physical scale \(R\),
\begin{equation}
\sigma^2(R) \equiv \left\langle \delta_R^2 \right\rangle ,
\end{equation}
and use it as a proxy for the expected relative size of NLO corrections. In this approach, values of \(\sigma^2(R) = \{0.1, 0.2, \ldots\}\) correspond to scales at which the NLO covariance is expected to contribute at the \(\{10\%, 20\%, \ldots\}\) level relative to the Gaussian term.

We define the smoothed density contrast \(\delta_R\) via a real-space spherical top-hat filter of radius \(R\),
\begin{equation}
\delta_R(\mathbf{x}) \equiv \frac{1}{V_R} \int_{|\mathbf{y}|\le R} d^3y \,
\delta_{\rm lin}(\mathbf{x}+\mathbf{y}),
\qquad
V_R = \frac{4\pi R^3}{3}.
\end{equation}
For a statistically homogeneous and isotropic density field, the corresponding variance may be written in Fourier space as
\begin{equation}
\label{eq:sigma2_general}
\sigma^2(R)
= \int \frac{d^3k}{(2\pi)^3}\,
P_{\rm lin}(k)\, W^2_{\rm TH}(kR) ,
\end{equation}
where \(P_{\rm lin}(k)\) is the linear matter power spectrum and
\begin{equation}
W_{\rm TH}(kR) = 3\,\frac{j_1(kR)}{kR}
\end{equation}
is the Fourier transform of the real-space top-hat window, with \(j_1\) the spherical Bessel function of order one.

Using isotropy to reduce Eq.~\eqref{eq:sigma2_general} to a one-dimensional integral, we obtain
\begin{equation}
\label{eq:sigma2_final}
\sigma^2(R)
= \frac{9}{2\pi^2}
\int_0^\infty dk\, k^2\,
P_{\rm lin}(k)\,
\left(\frac{j_1(kR)}{kR}\right)^2 .
\end{equation}

\begin{figure}[h!]
    \centering
    \includegraphics[width=0.8\linewidth]{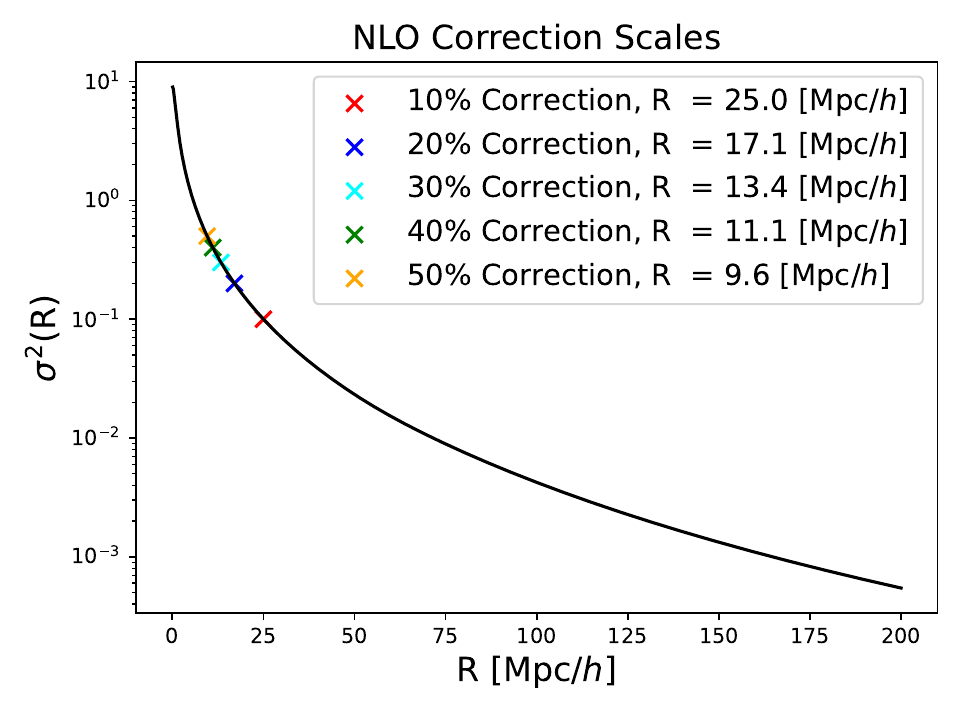}
    \caption{Estimate of the physical scales at which NLO corrections to the 4PCF covariance are expected to become significant. Shown is the variance of the linear density field smoothed to different scales $R$, $\sigma^2(R)=\langle \delta_R^2\rangle$; see Eq.~(\ref{eq:sigma2_final}). The legend indicates the $10\%$--$50\%$ correction levels, together with their corresponding scales and the markers used in the figure.}
    \label{fig:NLO_Correction_Scales}
\end{figure}

We evaluate Eq.~\eqref{eq:sigma2_final} numerically using the same linear power spectrum employed throughout this work. By solving for the values of \(R\) at which \(\sigma^2(R)\) reaches fixed thresholds, we identify the characteristic scales at which the NLO covariance is expected to induce fractional corrections of \(10\%\), \(20\%\), \(30\%\), \(40\%\), and \(50\%\) relative to the Gaussian covariance. The resulting scales are shown in Fig.~\ref{fig:NLO_Correction_Scales}.

We emphasize that this procedure provides an order-of-magnitude estimate rather than a sharp transition scale. Nonetheless, it offers a physically motivated and transparent criterion for assessing the regime of validity of the Gaussian covariance approximation for the 4PCF, and for identifying the scales on which NLO corrections are expected to become non-negligible.

\section{Computational Cost: Brute Force vs. Analytical Reduction}\label{sec:Computational_Cost}
In this section, we estimate the computational cost of evaluating the next-to-leading order second-order covariance of the 4PCF using a direct brute-force approach, and contrast it with the cost of our analytical reduction. The goal is to show that a naive numerical evaluation of the covariance expression is computationally prohibitive, while the reduced formulation is tractable in practice.

We begin by considering a brute-force (BF) example (without loss of generality) using the primary-primary configuration shown in Fig. \ref{fig:dc_configurations}, schematically written in configuration space as:
\begin{align}
{\rm BF} \propto\bigg< \int \frac{d^3\vecs}{V}\; \delta^{(2)}(\vecx + \vecr_{0})\delta^{(2)}(\vecx+\vecr'_{0}+\vecs)  \prod_{p=1}^{3} \prod_{t=1}^{3} \delta_{\rm lin.}(\vecx+\vecr_p)\delta_{\rm lin.}(\vecx+\vecr'_t+ \vecs) \bigg>.
\end{align}
Expressing the second-order density fields in Fourier space and expanding all linear density fields, the above expression becomes:
\begin{align}
&{\rm BF} \propto \bigg< \int_{\vecs}\;\int_{\veck_0}e^{-i\veck_{0}\cdot(\vecx+\vecr_0)}\int d^3\vecq_1\;d^3\vecq_2 \;\dD(\veck_0-\vecq_1-\vecq_2)\; F^{[2]}(\vecq_1,\vecq_2) \;\dlin(\vecq_1)\dlin(\vecq_2) \nonumber \\
& \qquad  \quad \times\int_{\veck'_0}e^{-i\veck'_{0}\cdot(\vecx+\vecr'_0+\vecs)} \int d^3\vecq'_1\;d^3\vecq'_2 \;\dD(\veck'_0-\vecq'_1-\vecq'_2) \;F^{[2]}(\vecq'_1,\vecq'_2) \;\dlin(\vecq'_1)\dlin(\vecq'_2) \nonumber \\
& \qquad\quad \times \prod_{p} \prod_{t}\int_{\veck_p}e^{-i\veck_{p}\cdot(\vecx+\vecr_p)} \dlin(\veck_p) \int_{\veck'_{t}}e^{-i\veck'_{t}\cdot(\vecx+\vecr'_p+\vecs)} \dlin(\veck'_t) \bigg>.
\end{align}
After performing the ensemble average using Wick’s theorem and carrying out the trivial integrations, the expression reduces to:
\begin{align}
&{\rm BF} \propto \int_{\vecs} \int d^3\vecq_1\;d^3\vecq_2 \;e^{-i(\vecq_1+\vecq_2)\cdot(\vecr_0-\vecr'_0-\vecs)} \;\left(F^{[2]}(\vecq_1,\vecq_2)\right)^2\;\Pk(q_1)\Pk(q_2)\nonumber \\
& \qquad \times  \prod_{p} \int_{\veck_p} \;e^{-i(\veck_p)\cdot(\vecr_p-\vecr'_p-\vecs)} \Pk(k_p). 
\end{align}

The integral over $\vecs$ can be performed directly, yielding a Dirac delta function that imposes the constraint $\veck_1+\veck_2+\veck_3+\vecq_1+\vecq_2=0$. This delta function may then be used to eliminate one of the four wave vectors, for example $\veck_3$. At this stage, the brute-force computation involves four independent three-dimensional momentum integrals (two \(\vecq\) integrals and two \(\veck_p\) integrals). Discretizing each dimension with \(N\) sampling points implies a total computational cost scaling as:
\begin{equation}
{\rm Cost}_{\rm BF} \sim N^{12}, \nonumber
\end{equation}
which is entirely infeasible for any realistic choice of \(N\). In contrast, our analytical reduction exploits translational invariance, angular-momentum decompositions, and the separability of radial and angular degrees of freedom to rewrite the same covariance contribution in terms of six nested one-dimensional radial integrals. In this case, the computational cost scales instead as:
\begin{equation}
{\rm Cost}_{\rm Reduced} \sim N^{6}, \nonumber
\end{equation}
representing a dramatic reduction in computational complexity. This reduction makes the numerical evaluation of the next-to-leading order 4PCF covariance practical, and is essential for any application to modern large-scale structure data.

\section{Convergence of the Projected Covariance}\label{sec:Convergence_Test}

In order to assess the convergence of the covariance contributions, we project the full covariance onto the isotropic basis by multiplying the results presented in $\S$\ref{sec:Resulting_Cov} by $\mathcal{P}^*_{\Omega}(\hatr_0,\hatr_1,\hatr_2,\hatr_3)$ and $\mathcal{P}^*_{\Omega'}(\hatr'_0,\hatr'_1,\hatr'_2,\hatr'_3)$, and integrating over all angular variables. This removes all angular dependence and allows us to study convergence for specific \emph{external} angular momentum configurations. Projecting the covariance onto this basis yields a scalar quantity for each choice of \emph{external} angular momenta
\begin{align}
\mathrm{Cov}_{\Lambda,\Lambda'; \;(T)}(R,R'), \nonumber
\end{align}
where we define the \emph{external} angular momenta $\Lambda \equiv \{L_0,L_1,L_2,L_3\}$ and $\Lambda' \equiv \{L'_0,L'_1,L'_2,L'_3\}$, $R \equiv \{r_0,r_1,r_2,r_3\}$ and $R' \equiv \{r'_0,r'_1,r'_2,r'_3\}$, and $T \in \{\mathrm{P\text{-}S}, \mathrm{H\text{-}H}, \mathrm{P\text{-}P}, \mathrm{PC\text{-}P\text{-}S}, \mathrm{PC\text{-}H\text{-}H}\}$ labels the covariance contribution.

To study convergence, we introduce an internal angular momentum cutoff $L_{\rm cutoff}$, restricting all \emph{internal} angular momenta appearing in the covariance expressions to satisfy $\{L_{\rm int}\} \leq L_{\rm Cutoff}$. The projected covariance truncated at this cutoff is then defined as
\begin{align}
\mathrm{Cov}^{(\leq L_{\rm cutoff})}_{\Lambda,\Lambda'; \;(T)}(R,R')
=
\sum_{\{L_{\rm int}\} \leq L_{\rm cutoff}}
\mathcal{A}_{\{L\},{T}} \;
\mathcal{R}_{\{L\},{T}}(R,R'),
\end{align}
where $\mathcal{A}_{\{L\},{T}}$ collects all angular prefactors and $\mathcal{R}_{\{L\},{T}}$ denotes the corresponding radial integrals. Then, we quantify convergence by comparing the truncated result to a reference calculation performed at a sufficiently large cutoff $L_{\rm cutoff}=10$:
\begin{align}
R(L_{\rm cutoff})\big|_{\Lambda,\Lambda';\,R,R'}
\equiv
\frac{
\mathrm{Cov}^{(\leq L_{\rm cutoff})}_{\Lambda,\Lambda'}(R,R')
}{
\mathrm{Cov}^{(\leq 10)}_{\Lambda,\Lambda'}(R,R')
}.
\label{eq:R_Lcutoff_def}
\end{align}
A value $R(L_{\rm cutoff}) \to 1$ indicates convergence of the internal angular momentum sums.

\begin{figure}[h!]
    \centering
    \includegraphics[width=0.8 \linewidth]{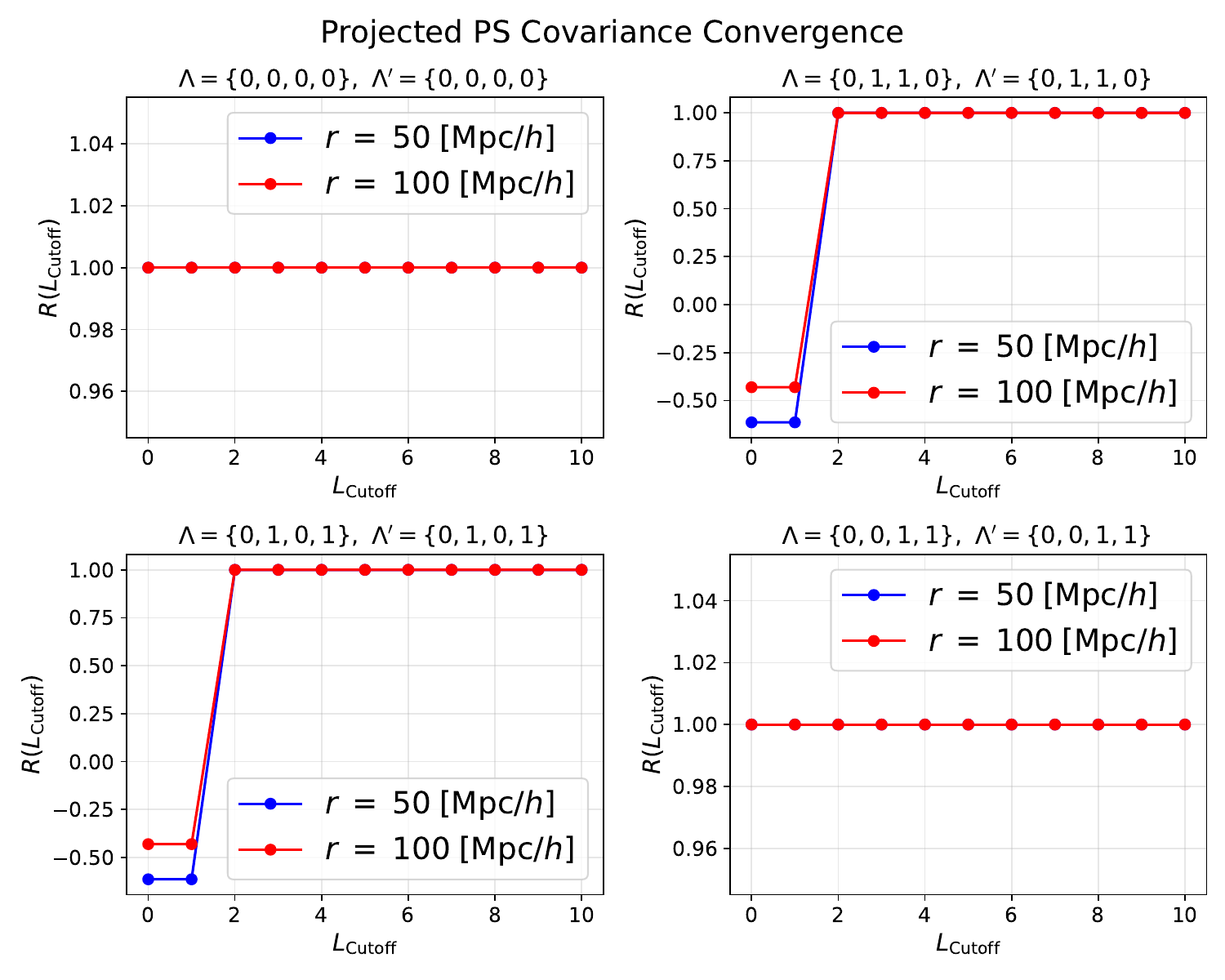}
    \caption{Convergence of the projected P--S covariance as a function of the internal angular momentum cutoff $L_{\rm Cutoff}$ for several fixed external configurations $(\Lambda,\Lambda')$. Each panel corresponds to a different choice of external angular momenta, while the lines show the ratio $R(L_{\rm cutoff})$ defined in Eq.~(\ref{eq:R_Lcutoff_def}) for the diagonal radial configurations $r_i = r'_i \in \{50,100\}\,[\mathrm{Mpc}/h]$ and $r_0 = r'_0 = 0$. The ratio approaches unity as $L_{\rm cutoff}$ increases, indicating convergence of the internal angular momentum sums. In all cases shown, the covariance is well converged by $L_{\rm cutoff} =2$.}
    \label{fig:Covaraince_Convergence}
\end{figure}

In this work, we focus on diagonal radial configurations with $r_1 = r_2 = r_3 = r'_1 = r'_2 = r'_3 = r \in \{50,100\}\,[\mathrm{Mpc}/h]$, and fix $r_0 = r'_0 = 0$. That is, we evaluate the covariance separately for $r = 50\,[\mathrm{Mpc}/h]$ and $r = 100\,[\mathrm{Mpc}/h]$. We remind the reader that $\vecr_0$ and $\vecr'_0$ were introduced to symmetrize the results presented in $\S$\ref{sec:Resulting_Cov}; consequently, they are set to zero when evaluating the covariance. As an example, Figure~\ref{fig:Covaraince_Convergence} shows the convergence of the projected P--S covariance for several representative external configurations of $(\Lambda,\Lambda')$. Each panel corresponds to a fixed choice of external angular momenta, while the horizontal axis denotes the internal cutoff $L_{\rm cutoff}$. As shown in the figure, the ratio $R(L_{\rm cutoff})$ rapidly approaches unity as the cutoff increases, indicating that only a small number of internal angular momenta contribute significantly to the covariance. In particular, convergence is typically achieved by $L_{\rm cutoff} = 2$, demonstrating that the covariance is well converged by $L_{\rm cutoff}=10$ for all configurations considered.

\end{document}